\documentclass[a4paper,fleqn]{cas-sc}

\usepackage[numbers]{natbib}
\usepackage{amsmath}
\usepackage{bm}
\usepackage{braket}
\usepackage{quantikz}
\usepackage{subfigure}
\usepackage[switch]{lineno}

\newcommand\etal{\mbox{\textit{et al.}}}
\def\tsc#1{\csdef{#1}{\textsc{\lowercase{#1}}\xspace}}
\tsc{WGM}
\tsc{QE}

\begin{document}
\let\WriteBookmarks\relax
\def\floatpagepagefraction{1}
\def\textpagefraction{.001}

\title [mode = title]{Quantum lattice Boltzmann method via density-matrix encoding for fluid simulation with wall boundary conditions}



%
\shortauthors{H. Su et~al.}
\shorttitle{}

\author[1]{Hao Su}

\author[2]{Boyuan Wang}


\affiliation[1]{organization={State Key Laboratory for Turbulence and Complex Systems, School of Mechanics and Engineering Science, Peking University},
            city={Beijing},
            postcode={100871}, 
            country={China}
            }

\author[1,2]{Yue Yang}[orcid=0000-0001-9969-7431]

\ead{yyg@pku.edu.cn}
\cormark[1]


\affiliation[2]{organization={HEDPS-CAPT, Peking University},
            city={Beijing},
            postcode={100871}, 
            country={China}
            }

\cortext[1]{Corresponding author}

\begin{abstract}
The quantum lattice Boltzmann method (QLBM) holds promise for efficient fluid simulations, yet the treatment of realistic boundary conditions, particularly solid wall boundary with arbitrary geometry, remains a critical open challenge. We propose a QLBM algorithm that incorporates wall boundaries as well as inlet/outlet velocity conditions. Building upon the existing QLBM with ensemble transformations, we adopt a density-matrix encoding that offers greater flexibility in circuit design and enables the implementation of non-unitary operations through Kraus operators. For boundary enforcement, we design a quantum implementation of the half-way bounce-back scheme and introduce a component exchange step prior to the streaming step, which naturally enforces the no-slip condition without modifying the transport operation. Inlet and outlet conditions are imposed via controlled SWAP operations combined with ancilla registers prepared in the prescribed velocity states. Our QLBM is quantitatively validated through simulations of two-dimensional flows past a backward-facing step, a cylinder, and an obstacle with complex geometry, demonstrating both accuracy and extensibility of our boundary algorithm. 
\end{abstract}

\begin{highlights}
\item Density-matrix encoding for flexible boundary operations in the quantum lattice Boltzmann method
\item Component exchange step enforcing arbitrary-geometry no-slip walls without altering streaming
\item Validation on the flows past a backward-facing step, a cylinder, and an obstacle with complex geometry 
\end{highlights}


\maketitle

\section{Introduction}\label{sec:introduction}
Computational fluid dynamics (CFD) has long been dominated by classical numerical methods for solving the Navier-Stokes (NS) equations and related partial differential equations. As engineering and scientific demands for simulation accuracy and scale continue to escalate, traditional CFD approaches face increasingly severe computational bottlenecks. High-fidelity simulations, such as direct numerical simulation (DNS), require immense grid resolutions and extremely small time steps in complex geometries and high-Reynolds-number flows, leading to computational complexity of $O(\mathrm{Re}^3)$~\cite{Pope2000}, where $\mathrm{Re}$ denotes the Reynolds number. These challenges have motivated the search for novel computing paradigms. Quantum computing~\cite{Feynman1982, Nielsen2010}, leveraging the principles of superposition and entanglement inherent in quantum mechanics, has demonstrated significant advantages over classical methods in certain specific problems~\cite{Shor1999, Grover1996}. The exponential memory efficiency and quantum parallelism of the quantum computing paradigm make it especially attractive for the increasingly demanding large-scale CFD applications of today. Consequently, quantum computing of fluid dynamics (QCFD) has emerged as a promising alternative to classical CFD.

A fundamental distinction lies in the fact that quantum computing operates under unitary and reversible transformations, whereas the NS equations of classical CFD are inherently non-linear and irreversible~\cite{Succi2023, Bharadwaj2025}. This mismatch compels QCFD solvers to reformulate the governing equations through multiple techniques to preserve quantum compatibility. Current QCFD research encompasses several distinct paradigms, including the Carleman linearization~\cite{Liu2021, Sanavio2024}, the hydrodynamic Schr\"odinger equation~\cite{Meng2024}, the Koopman method~\cite{Zhang2025}, the quantum homotopy analysis~\cite{Xue2025}, the linear combination of Hamiltonian simulation~\cite{Meng2026} and so on. In particular, the quantum lattice Boltzmann method (QLBM), rooted in the classical LBM framework~\cite{Kruger2016, Chen1998}, has emerged as one of the most promising QCFD approaches. By viewing the fluid as an ensemble of mesoscopic particles, the classical LBM framework reformulates the NS equations into a higher-dimensional lattice gas representation, which naturally converts the nonlinear convection term into a linear transport term. The particle distribution functions evolve through local collision and streaming steps, offering exceptional parallelism and geometric flexibility that are inherently compatible with quantum computing. 

Currently, the practical deployment of QLBM is confronted with two fundamental challenges: the nonlinear collision term, such as the Bhatnagar-Gross-Krook (BGK) model~\cite{Qian1992}, and the difficulty of imposing realistic boundary conditions. To address the former issue, two principal strategies have been pursued in the literature. One strategy is to neglect the non-linear contribution of collision term~\cite{Ljubomir2022, Wawrzyniak2025, Kocherla2024, Xu2025, Zeng2025} or to eliminate the nonlinearity through techniques like Carleman linearization~\cite{Itani2022, Itani2024, Jennings2025, Zamora2026}. This strategy either confines the applicable problem domain or incurs an exponential growth in the system dimension. The other strategy resorts to lattice gas cellular automata (LGCA) model~\cite{Hardy1973, Frisch2019}, which contains physical collision rules that are intrinsically linear and sidestep the nonlinearity problem. Yet its quantum algorithm suffers from extremely high dimensionality~\cite{Wolf2004}. The two approaches each adopt distinct boundary treatment strategies: the first incorporates boundary conditions directly into the linearized operator~\cite{Bakker2023, Turro2025, Kumar2025, Yang2026, Jennings2026}, while the second introduces additional collision rules between physical particles and the boundary during the streaming and collision steps~\cite{Schalkers2024, Georgescu2025, Georgescu2026}. 

Combining the moderate dimensionality of LBM with the linear collision treatment of LGCA, Wang \etal~\cite{Wang2025} proposed a QLBM framework featuring a novel ensemble description of lattice gas compatible with quantum computing, together with an additional H-step to maintain the particle ensemble near the thermodynamic equilibrium. However, this study was restricted to periodic boundary conditions, and incorporating more general boundary conditions is challenging due to the unitary constraints of quantum computing. 

\begin{figure}
    \centering
    \includegraphics[width=\linewidth]{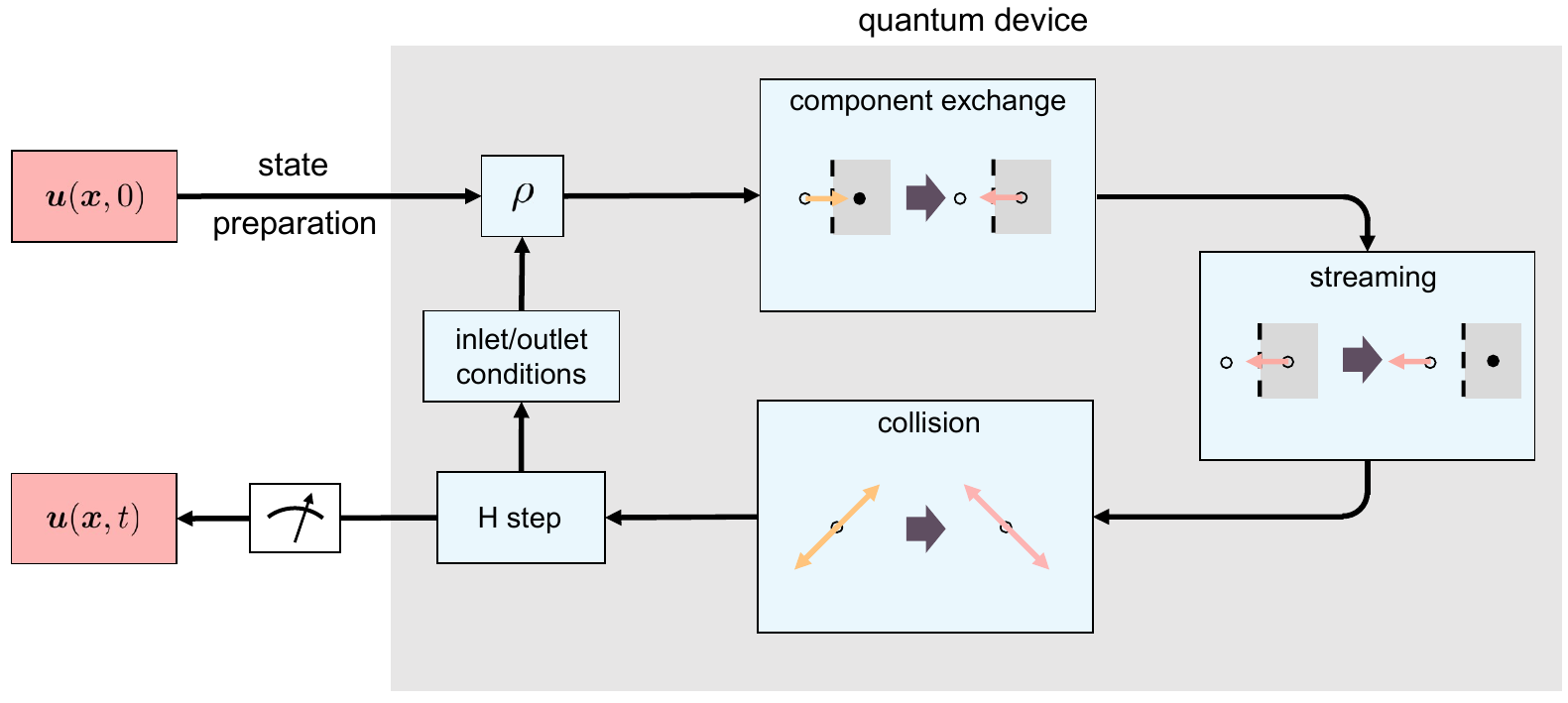}
    \caption{The overall procedure for our QLBM with general boundary scheme. The initial velocity field $\bm{u}(\bm{x}, 0)$ is encoded into a general mixed state. The quantum state then undergoes three consecutive operations: component exchange, streaming, and collision. Together, these operations are equivalent to classical streaming and collision steps with wall boundary treatment. An additional H-step is subsequently applied to drive the quantum state toward thermodynamic equilibrium. Finally, the inlet and outlet conditions are imposed. This sequence of steps are iterated until the target simulation time is reached, after which the final results are extracted via quantum measurement.}
    \label{fig:procedure}
\end{figure}

To overcome this limitation, we encode the probability distribution of the ensemble with a density-matrix formulation in this study, which offers greater flexibility in circuit design compared to pure state encoding. Building on this encoding, we extend the QLBM~\cite{Wang2025} by incorporating wall boundary conditions with general geometries. By inserting the component exchange step immediately before the streaming step, the wall boundary is naturally satisfied without modification of streaming step. Furthermore, the inlet and outlet conditions are implemented by introducing extra ancilla qubits under the new encoding. The overall procedure for our QLBM is illustrated in Figure~\ref{fig:procedure}. 
The extended QLBM is validated on several 2D benchmark flows, including decaying Poiseuille flow, backward-facing step flow, and flow past a circular cylinder. To further demonstrate the potential of our boundary treatment, we also test our algorithm on a flow past a obstacle with more complex geometry, confirming its applicability beyond canonical test cases.

The remainder of this paper is organized as follows. Section~\ref{sec:QLBM} briefly introduces the QLBM, together with our extension with the density-matrix encoding. Section~\ref{sec:boundary} presents our boundary treatment algorithm for wall boundary and inlet/outlet conditions, along with the overall complexity analysis. Section~\ref{sec:result} assesses our algorithm with the benchmark flows. Section~\ref{sec:conclusion} concludes the paper with a discussion of our algorithm.

\section{QLBM with density-matrix encoding}\label{sec:QLBM}
\subsection{Review of the existing QLBM}
We first review the existing QLBM procedure. 
The Boltzmann equation describes fluid with a distribution function $f(\bm{x}, \bm{v}, t)$. This function denotes the probability density of finding a particle with velocity $\bm{v}$ at time $t$ and position $\bm{x}$ in a $d$-dimensional space, where $d$ is the spatial dimension. The Boltzmann equation takes the form
\begin{equation}\label{eq:BE}
    \dfrac{\partial f}{\partial t}+\bm{v}\cdot\nabla_x f=\Omega(f),
\end{equation}
where $\bm{v}\cdot\nabla_x f$ is the advection term, and $\Omega(f)$ describes local collisions between particles. Macroscopic quantities, such as fluid density $\rho$ and momentum $\rho\bm{u}$, are related to $f$ by
\begin{equation}
    \rho(\bm{x}, t)=\int f(\bm{x}, \bm{v}, t)\mathrm{d}\bm{v},\qquad \rho(\bm{x},t)\bm{u}(\bm{x},t)=\int f(\bm{x},\bm{v},t)\bm{v}\mathrm{d}\bm{v}.
\end{equation}

\begin{figure}
    \centering
    \includegraphics[width=0.65\linewidth]{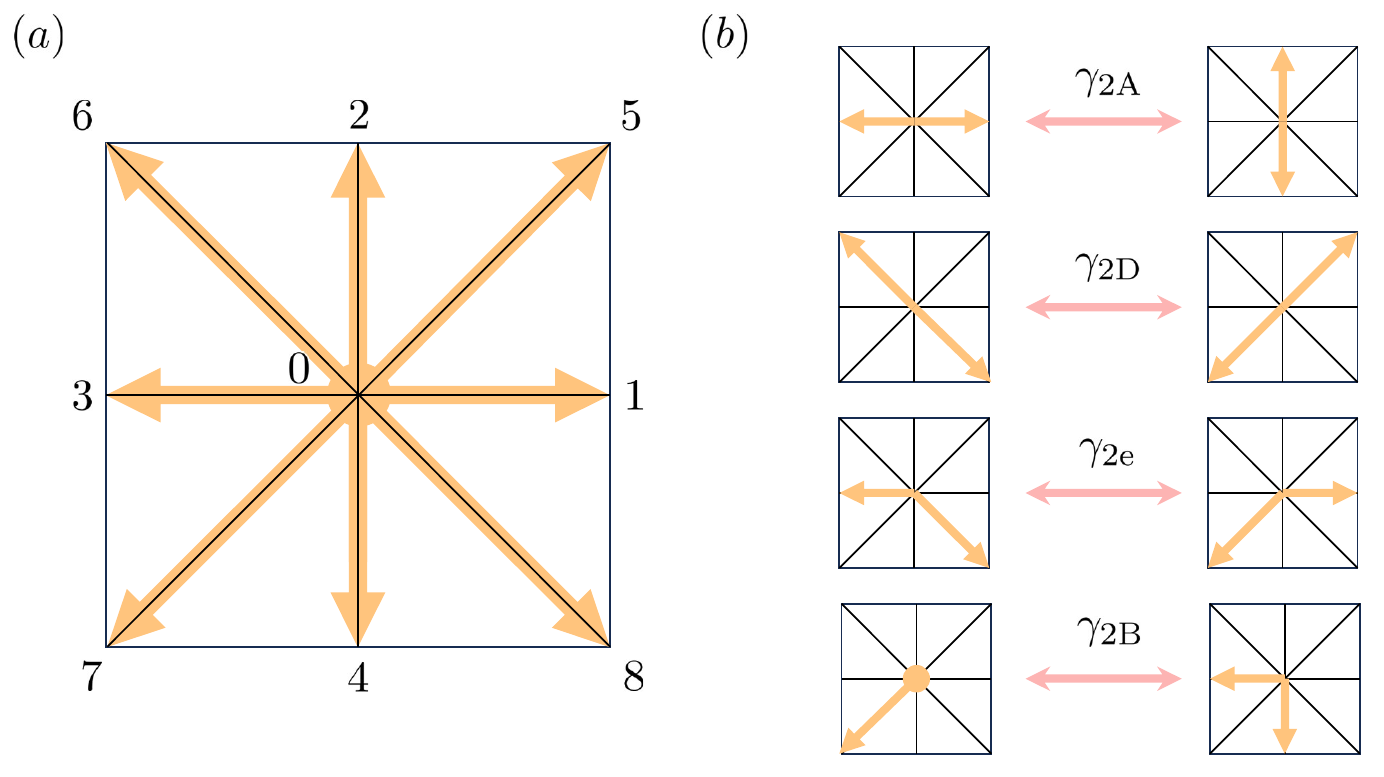}
    \caption{(a) Examples of 2D discrete velocity sets, with each velocity direction labeled by an index number. Indices 1 through 4 constitute the D2Q4 model, while indices 0 through 8 form the D2Q9 model. (b) Linear collision rules for the D2Q9 model. Each rule is labeled with its probability above the arrow. The first collision rule also serves as the collision rule for the D2Q4 model. }
    \label{fig:collision_model}
\end{figure}

Equation~\eqref{eq:BE} is difficult to solve directly due to its high dimensionality, therefore requires further discretization to achieve efficient numerical simulation. Two major discretization schemes for the Boltzmann equation are the LGCA and LBM. Both schemes replace the continuous velocity space with $q$ discrete velocity values $\bm{e}_m$, $m=1, \cdots, q$, which is referred to as the D$d$Q$q$ model. Figure \ref{fig:collision_model}(a) shows examples of common 2D discrete velocity sets. The D2Q4 model includes the velocity directions labeled 1 through 4, while the D2Q9 model includes directions labeled 0 through 8. The distribution function is thus represented as a vector-valued function $\bm{f}(\bm{x}, t)$, where each component $f_m(\bm{x}, t)$ denotes the probability density of finding a particle with velocity $\bm{e}_m$. Macroscopic quantities can be acquired similarly as
\begin{equation}
    \rho(\bm{x}, t)=\sum_m f_m(\bm{x}, t),\qquad \rho(\bm{x}, t)\bm{u}(\bm{x}, t)=\sum_m\bm{e}_mf_m(\bm{x}, t).
\end{equation}

The main difference between LBM and LGCA is the treatment of $\Omega$ and the value range of $\bm{f}$. In the LGCA, the kinetic evolution is described with physical particles following a set of linear collision rules. Figure~\ref{fig:collision_model}(b) illustrates the set of collision rules for the D2Q9 model, with $\gamma$ denoting the collision probability for each rule. Note that the last two rules can occur in four distinct directions. The first rule alone constitutes the collision rule for the D2Q4 model. The components of $\bm{f}$ take only the values 0 and 1. In contrast, the LBM models the collision as a relaxation process toward a local equilibrium state, and components of $\bm{f}$ take continuous real values instead of Boolean values. However, both approaches are difficult to implement directly on quantum devices. The high dimensionality of LGCA offsets the speedup gained from quantum computing, while LBM contains nonlinear collision terms that are incompatible with inherently linear operations of quantum hardware. 

Wang \emph{et al.}~\cite{Wang2025} proposed a meso-level framework integrating both methods, obtaining a QLBM ensemble with medium dimensionality and linear collision terms. 
In this QLBM, $\bm{f}$ on each node is described by all possible combinations of $q$ velocity values, denoted by $q$-bit binary string $\bm{n}=n_{q-1}\cdots n_1n_0$, where each $n_j$ takes Boolean value of $0$ or $1$ corresponding to the absence or presence of particles. For example, in the D2Q5 model, $\bm{n}=00011$ refers to the state where two particles with velocities $\bm{e}_1$ and $\bm{e}_2$ are occupying the node. On each node, each combination $\bm{n}$ is given a probability $p^{(\bm{n})}$ satisfying the normalization condition $\sum_{\bm{n}}p^{(\bm{n})}=1$. This provides a more detailed representation of the local particle distribution, where the velocity distribution function $\bm{f}=[f_0, \cdots, f_{q-1}]$ is related to $p^{(\bm{n})}$ by
\begin{equation}\label{eq:ptof}
    f_j=\sum_{\bm{n}}n_jp^{(\bm{n})},\quad 0\leq j\leq q-1.
\end{equation}
Different velocity values are assumed to be independently distributed, in accordance with the classical LBM. The $2^q$ combinations are introduced for linearity during simulation. 

For a lattice with $N$ nodes, different nodes are assumed to be independent of each other~\cite{Wang2025}, which achieves medium dimensionality of $2^qN$ for the entire construction. Each realization is described by $\{\bm{x}, \bm{n}\}$, corresponding to probability $p_{\bm{x}, \bm{n}}$ with the normalization condition $\sum_{\bm{x}, \bm{n}}p_{\bm{x}, \bm{n}}=1$.

At each time-marching step, the streaming step is realized as 
\begin{equation}\label{eq:transport}
    \{\bm{x}, \bm{n}\}\mapsto\sum_{j=0}^{q-1}\dfrac{n_j}{\rho_{\bm{n}}}\{\bm{x}+\bm{e}_j\Delta x,\bm{n}\},
\end{equation}
where $\rho_{\bm{n}}=\sum_{j=0}^{q-1}n_j$ is the number of particles at the node. During the streaming step, local realization is equally divided into $\rho_{\bm{n}}$ branches, and each branch propagates into neighboring node along direction $\bm{e}_j$ with $n_j=1$. The collision step takes place locally, following a set of rules predefined collision rules. It is realized as 
\begin{equation}\label{eq:collision}
    \{\bm{x}, \bm{n}\}\mapsto \gamma\{\bm{x}, \bm{n}^*\}+(1-\gamma)\{\bm{x}, \bm{n}\},
\end{equation}
where $\bm{n}^*$ denotes the particle combination after collision, and $\gamma\in[0,1]$ is the probability of collision. This probability affects the fluid viscosity through the relation~\cite{Chen1989}
\begin{equation}
    \mu=\dfrac{\epsilon}{\gamma_{\mathrm{2D}}\epsilon^2+\gamma_{\mathrm{2B}}(1-\epsilon)^2}\Delta t,
\end{equation}
where $\epsilon$ denotes the internal energy, $\Delta t$ denotes the single time step, and $\gamma_{\mathrm{2D}}$ and $\gamma_{\mathrm{2B}}$ are the probabilities associated with different collision rules as in Figure~\ref{fig:collision_model}(b). In this work, we set $\epsilon=1/3$ and $\gamma=0.5$ for all collision rules. 

After streaming and collision steps, the independence assumption for different velocity values within same node no longer holds, resulting in unphysical simulation results. To maintain the independence, the extra H-step is introduced to regulate detailed distribution $p^{(\bm{n})}$. Physically, this corresponds to the relaxation process to equilibrium state. During the H-step, the velocity distribution $f_j(\bm{x})$ is calculated via Equation~\eqref{eq:ptof}. This distribution is directly related to macroscopic quantities and should be kept constant during the H-step. After that, to achieve intra-node independence, the particle distribution is reformed as
\begin{equation}
    p^{(\bm{n})}=\prod_{n_j\neq 0}f_j\prod_{n_j=0}(1-f_j),
\end{equation}
which ensures constant velocity distribution and mass conservation while regains independence. 

\subsection{Quantum encoding with mixed states}

We extend the QLBM~\cite{Wang2025} by using the general mixed state $\varrho$ to encode probability $p_{\bm{x}, \bm{n}}$, thereby circumventing the restrictions imposed by purely unitary operations. 
This density-matrix encoding satisfies 
\begin{equation}\label{eq:rho_condition}
    \bra{\bm{x}_0}\bra{\bm{n}_0}\varrho\ket{\bm{n}_0}\ket{\bm{x}_0}=p_{\bm{x}_0, \bm{n}_0},\quad \forall \bm{x}_0, \bm{n}_0.
\end{equation}
Note that $\ket{\bm{n}}\ket{\bm{x}}$ form a orthogonal complete basis for Hilbert space $\mathcal{H}^{\otimes (q+n_g)}$, where $n_g=\lceil \log_2N\rceil$ is the number of qubits required to encode all grid points. In other words, the probability distribution of $\ket{\bm{x}}\ket{\bm{n}}$ is recorded on the diagonal elements of $\varrho$. Note that this encoding is not unique, where many different $\varrho$ can correspond to the same distribution. 

General QLBM operations, like the operations in Equations~\eqref{eq:transport} and \eqref{eq:collision}, can be described by stochastic matrices $P_{ij}\in\mathbb{R}^{2^n\times2^n}$ acting on diagonal elements of $\varrho$. Such matrices satisfy $P_{ij}\geq 0$ and $\sum_iP_{ij}=1$, ensuring the probability is preserved. However, these matrices are not guaranteed to be unitary, which makes them difficult to implement directly on a quantum device using pure-state encoding. Nevertheless, they can be included under the broader framework of Kraus operators~\cite{Kraus1971}, which is a generalization of unitary maps in open quantum systems. According to the Kraus theorem, a linear map $\mathcal{E}(\varrho)$ is completely positive and trace-preserving (CP-TP) map if and only if it can be written in the form
\begin{equation}\label{eq:Kraus_definition}
    \mathcal{E}(\varrho)=\sum_{\alpha=1}^{D}K_{\alpha}\varrho K_{\alpha}^{\dagger},
\end{equation}
where the Kraus operators $K_{\alpha}$ are linear and satisfy
\begin{equation}\label{eq:Kraus_condition}
    \sum_{\alpha=1}^{D}K_{\alpha}^{\dagger}K_{\alpha}=I.
\end{equation}
A map $\mathcal{E}$ is said to be positive if it maps any positive operator to a positive operator, whereas complete positivity means that any extension $\mathcal{E}\otimes I_m$ with $m\ge 0$ remains a positive map. The trace preserving means that $\mathrm{Tr}(\mathcal{E}\varrho)=\mathrm{Tr}(\varrho)$ for every density matrix $\varrho$.

QLBM operations should satisfy the mathematical conditions of $\mathcal{E}$. The completely positiveness is straightforward from the non-negativity constraint of $P_{ij}$ elements, while the trace preserving corresponds to the condition $\sum_iP_{ij}=1$. Thus, according to the Kraus theorem, any stochastic matrix admits a Kraus operator representation, which can be explicitly constructed as
\begin{equation}\label{eq:Kraus_construction}
    K_{ij}=\sqrt{P_{ij}}\ket{x_i}\bra{x_j},
\end{equation}
where $\ket{x_i}$ refers to general complete orthogonal basis for given Hilbert space. In the QLBM, this basis is precisely $\ket{\bm{x}}\ket{\bm{n}}$. The non-negativity of $P_{ij}$ ensures the validity of square root operations. It is straightforward to verify that this construction satisfies condition in Equation~\eqref{eq:Kraus_condition} as
\[\sum_{i,j}K_{ij}^\dagger K_{ij}=\sum_{i,j}P_{ij}\ket{x_j}\bra{x_j}=\sum_j(\sum_i P_{ij})\ket{x_j}\bra{x_j}=\sum_j\ket{x_j}\bra{x_j}=I.\]
The equivalence between the constructed Kraus operators and original stochastic matrix can be verified by direct calculation
\[\begin{aligned}
\bra{x_k}\tilde{\varrho}\ket{x_k}&=\sum_{i,j}\bra{x_k}K_{ij}\varrho K_{ij}^\dagger\ket{x_k}\\
&=\sum_{i,j}P_{ij}\braket{x_k|x_i}\bra{x_j}\varrho\ket{x_j}\braket{x_i|x_k}\\
&=\sum_{i,j}P_{ij}\delta_{ik}^2\bra{x_j}\varrho\ket{x_j}\\
&=\sum_{j}P_{kj}\bra{x_j}\varrho\ket{x_j},
\end{aligned}\]
showing that the diagonal elements are transformed by the stochastic matrix $P_{ij}$.

The fact that any stochastic matrix corresponds to a CP-TP map is highly advantageous for the implementation on quantum devices, as the Stinespring dilation theorem~\cite{Choi1975} states that any CP-TP map can be represented as
\[\varrho\rightarrow \mathrm{Tr}_E[U(\varrho\otimes E)U^\dagger].\]
This implies that under density-matrix encoding, all CP-TP maps can be implemented by introducing several ancilla qubits and tracing them off after certain unitary operations. This result ensures the feasibility of our QLBM operations on real quantum device. 

\subsection{Quantum circuit implementation}
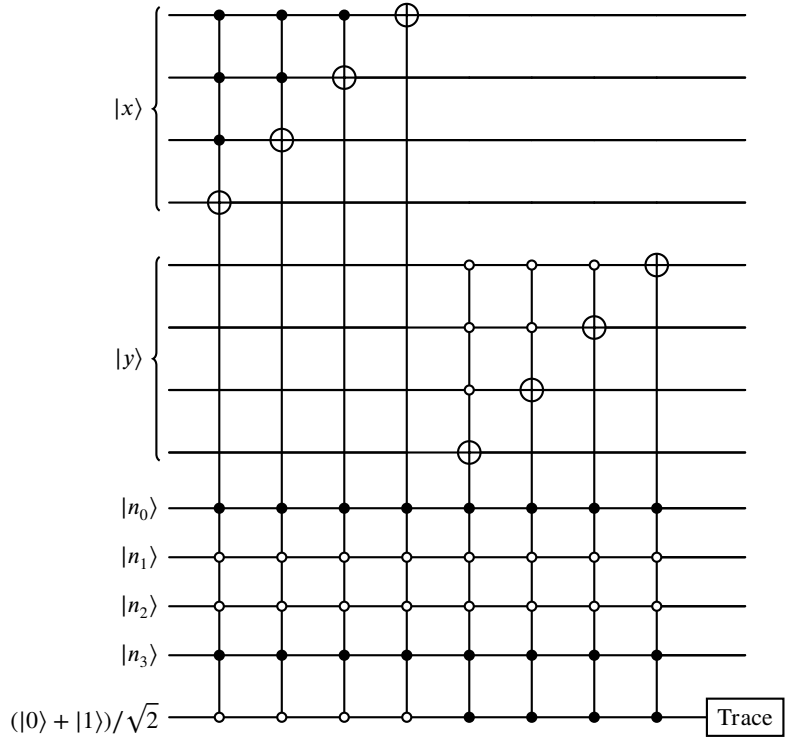
\begin{figure}
    \centering
    \begin{quantikz}
        \lstick[wires=4]{$\ket{x}$} &\ctrl{1} &\ctrl{1} &\ctrl{1} &\targ{} &\qw &\qw &\qw &\qw &\qw \\
        &\ctrl{1} &\ctrl{1} &\targ{} &\qw &\qw &\qw &\qw &\qw &\qw \\
        &\ctrl{1} &\targ{} &\qw &\qw &\qw &\qw &\qw &\qw &\qw \\
        &\targ{} &\qw &\qw &\qw &\qw &\qw &\qw &\qw &\qw \\
        \lstick[wires=4]{$\ket{y}$} &\qw &\qw &\qw &\qw &\octrl{1} &\octrl{1} &\octrl{1} &\targ{} &\qw \\
        &\qw &\qw &\qw &\qw &\octrl{1} &\octrl{1} &\targ{} &\qw &\qw \\
        &\qw &\qw &\qw &\qw &\octrl{1} &\targ{} &\qw &\qw &\qw \\
        &\qw &\qw &\qw &\qw &\targ{} &\qw &\qw &\qw &\qw \\
        \lstick{$\ket{n_0}$} &\ctrl{-5} &\ctrl{-6} &\ctrl{-7} &\ctrl{-8} &\ctrl{-1} &\ctrl{-2} &\ctrl{-3} &\ctrl{-4} &\qw \\
        \lstick{$\ket{n_1}$} &\octrl{-1} &\octrl{-1} &\octrl{-1} &\octrl{-1} &\octrl{-1} &\octrl{-1} &\octrl{-1} &\octrl{-1} &\qw \\
        \lstick{$\ket{n_2}$} &\octrl{-1} &\octrl{-1} &\octrl{-1} &\octrl{-1} &\octrl{-1} &\octrl{-1} &\octrl{-1} &\octrl{-1} &\qw \\
        \lstick{$\ket{n_3}$} &\ctrl{-1} &\ctrl{-1} &\ctrl{-1} &\ctrl{-1} &\ctrl{-1} &\ctrl{-1} &\ctrl{-1} &\ctrl{-1} &\qw \\
        \lstick{$(\ket{0}+\ket{1})/\sqrt{2}$} &\octrl{-1} &\octrl{-1} &\octrl{-1} &\octrl{-1} &\ctrl{-1} &\ctrl{-1} &\ctrl{-1} &\ctrl{-1} &\gate{\mathrm{Trace}}
    \end{quantikz}
    \caption{Quantum circuit for the streaming step for $\bm{n}=1001$ under 2D $16\times 16$ grid and D2Q4 model. The extra qubit is initialized as $(\ket{0}+\ket{1})/\sqrt{2}$ and traced out after the process. Cyclic permutation $\ket{x}\rightarrow\ket{x+1}$ is performed on $\ket{x}$ coordinate part, and $\ket{y}\rightarrow\ket{y-1}$ is performed on $\ket{y}$ coordinate part.}
    \label{fig:transport}
\end{figure}

We now describe the quantum circuit implementation of the streaming and collision steps in Equations~\eqref{eq:transport} and \eqref{eq:collision} with the density-matrix encoding. The circuits described in this section are essentially the same as those in Ref.~\cite{Wang2025}, however, we reinterpret them here within the more general density matrix framework. 
The streaming step can be realized by the circuit shown in Figure~\ref{fig:transport}. The quantum state $\varrho$ is equally partitioned into $\rho_{\bm{n}}$ branches using $\lceil \log_2\rho_{\bm{n}}\rceil$ ancilla qubits, and the coordinate part of each branch is shifted according to the nonzero bits of the binary string $\bm{n}$. As an example, we illustrate a 2D case in Figure~\ref{fig:transport}, where the coordinate qubit register $\ket{\bm{x}}$ is further split into two dimensions $\ket{y}\ket{x}$. The coordinate shift is implemented by the quantum incrementer and decrementer~\cite{Li2014}, which consists of $O(n_g)$ controlled X gates. Note that the periodic boundary condition is assumed here, as the incrementer and decrementer essentially perform a cyclic permutation. For a pure state input $\ket{n}\ket{y}\ket{x}$, the circuit in Figure~\ref{fig:transport} performs the transformation
\[\begin{aligned}
\ket{n}\ket{y}\ket{x}&\stackrel{\mathrm{ancilla}}\longrightarrow \dfrac{1}{\sqrt{2}}(\ket{0}+\ket{1})\ket{n}\ket{y}\ket{x}\\
&\longrightarrow \dfrac{1}{\sqrt{2}}\ket{0}\ket{n}\ket{y}\ket{x+1}+\dfrac{1}{\sqrt{2}}\ket{1}\ket{n}\ket{y-1}\ket{x}\\
&\stackrel{\mathrm{trace}}\longrightarrow\dfrac{1}{2}\ket{n}\ket{y}\ket{x+1}\bra{x+1}\bra{y}\bra{n}+\dfrac{1}{2}\ket{n}\ket{y-1}\ket{x}\bra{x}\bra{y-1}\bra{n}.
\end{aligned}\]
The circuit output is a mixed state consisting of an equally weighted mixture of the shifted states $\ket{n}\ket{y}\ket{x+1}$ and $\ket{n}\ket{y-1}\ket{x}$. This circuit transforms pure states into mixed states while satisfying the probability condition in Equation~\eqref{eq:rho_condition}.

\begin{figure}
    \subfigure[]{
        \begin{quantikz}
            \lstick{$\ket{n_0}$} &\ctrl{1} &\qw &\targ{} &\qw &\ctrl{2} &\targ{} &\ctrl{2} &\qw &\ctrl{1} &\qw\\
            \lstick{$\ket{n_1}$} &\targ{} &\ctrl{2} &\qw &\qw &\qw &\qw &\qw &\ctrl{2} &\targ{} &\qw\\
            \lstick{$\ket{n_2}$} &\ctrl{1} &\qw &\qw &\targ{} &\targ{} &\ctrl{-2} &\targ{} &\qw &\ctrl{1} &\qw\\
            \lstick{$\ket{n_3}$} &\targ{} &\ctrl{1} &\qw &\qw &\qw &\qw &\qw &\ctrl{1} &\targ{} &\qw\\
            \lstick{$\ket{0}_a$} &\qw &\targ{} &\ctrl{-4} &\ctrl{-2} &\ctrl{-2} &\ctrl{-2} &\ctrl{-2} &\targ{} &\qw &\qw
        \end{quantikz}
        \label{fig:collision_D2Q4}
    }
    \subfigure[]{
        \begin{quantikz}
            \lstick{$\ket{\bm{x}}$} &\qw &\qw \\
            \lstick{$\ket{\bm{n}}$} &\gate{\mathrm{Collision}} &\qw \\
            \lstick{$\sqrt{1-\gamma}\ket{0}+\sqrt{\gamma}\ket{1}$} &\ctrl{-1} &\gate{\mathrm{Trace}}
        \end{quantikz}
        \label{fig:collision_general}
    }
    \caption{(a) Quantum circuit for the collision step under D2Q4 model, which exchanges $\ket{0101}$ and $\ket{1010}$ and maintains other states. (b) General realization of collision probability $\gamma$ with one ancilla qubit. }
    \label{fig:collision}
\end{figure}
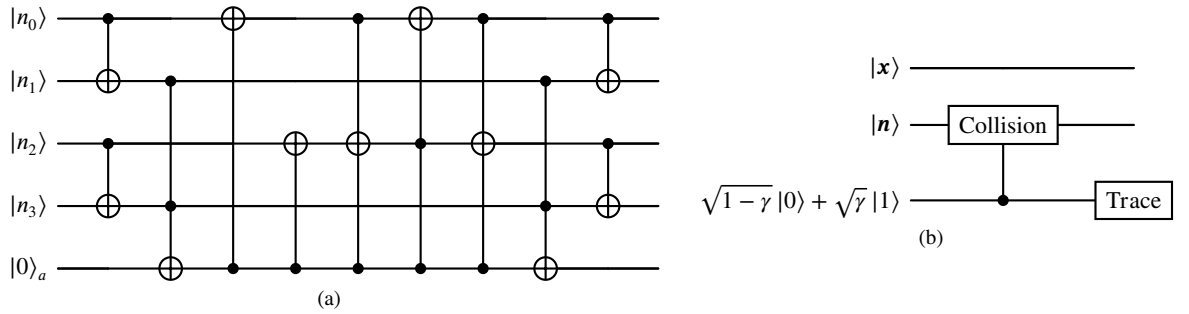

A general set of collision rules can be viewed as a specific permutation of the computational basis states $\bm{n}$, which corresponds to a unitary operation, and circuit construction for such permutations has already been proposed~\cite{Hanson2025}. As an example, quantum realization of the collision step in D2Q4 model is shown in Figure~\ref{fig:collision_D2Q4}, where the operation is concisely implemented using only a few CNOT and Toffoli gates. The collision probability $\gamma$ in Equation~\eqref{eq:collision} is realized by an ancilla qubit, as illustrated in Figure~\ref{fig:collision_general}. The ancilla qubit is initialized to $\sqrt{1-\gamma}\ket{0}+\sqrt{\gamma}\ket{1}$, and the collision step is applied conditioned on the ancilla being $\ket{1}$ state, which occurs with probability $\gamma$. For a pure input state $\ket{\bm{n}}\ket{\bm{x}}$, the circuit executes the transformation
\[\begin{aligned}
\ket{\bm{n}}\ket{\bm{x}}&\stackrel{\mathrm{ancilla}}\longrightarrow (\sqrt{1-\gamma}\ket{0}+\sqrt{\gamma}\ket{1})\ket{\bm{n}}\ket{\bm{x}}\\
&\longrightarrow\sqrt{1-\gamma}\ket{0}\ket{\bm{n}}\ket{\bm{x}}+\sqrt{\gamma}\ket{1}\ket{\bm{n^*}}\ket{\bm{x}}\\
&\stackrel{\mathrm{trace}}\longrightarrow (1-\gamma)\ket{\bm{n}}\ket{\bm{x}}\bra{\bm{x}}\bra{\bm{n}}+\gamma\ket{\bm{n^*}}\ket{\bm{x}}\bra{\bm{x}}\bra{\bm{n^*}},
\end{aligned}\]
where $\ket{\bm{n^*}}$ denotes the velocity configuration after collision. The circuit outputs a mixed state consisting of the collided state $\ket{\bm{n^*}}$ and the original state $\ket{\bm{n}}$ with probability $\gamma$ and $(1-\gamma)$. Similarly, this circuit turns a pure state into a mixed state. 

The H-step is mathematically equivalent to disentangling the qubits in the $\ket{\bm{n}}$ register, which can be realized by extracting $\ket{\bm{n}}$ qubits from different copies and recombining them with the same register $\ket{\bm{x}}$. The quantum implementation of this step is essentially the same as in Ref.~\cite{Wang2025}. 

\section{Boundary treatment in QLBM}\label{sec:boundary}

We propose the treatment of static solid wall boundary conditions and inlet/outlet conditions for the QLBM with the density-matrix encoding. By depicting the boundary shape with grid points, static wall boundary with complex geometry can be realized. In addition, the density-matrix encoding enables the implementation of inlet and outlet conditions using ancilla qubits, where the prescribed velocity profiles are imposed by appropriately preparing the ancilla state amplitudes. Together, these techniques extend the applicability of our QLBM scheme to a wide range of practical flow configurations. 

\subsection{Wall boundary}
For the static wall boundary condition, we employ the halfway bounce-back rule within a link-wise boundary scheme~\cite{Kruger2016}, i.e., the wall is located halfway between two grid points. Particles stream toward the wall and bounce back to the same grid point within a single time step, with their velocity direction reversed upon reflection. This scheme can be expressed as
\begin{equation}
    f_{-j}(\bm{x}_b, t+\Delta t)=f_j(\bm{x}_b,t),
\end{equation}
where subscript $-j$ denotes the direction opposite to $\bm{e}_j$, and $\bm{x}_b$ is a fluid grid point immediately adjacent to the boundary. This boundary treatment is numerically stable and second-order accurate. 

This boundary scheme can be extended to the moving wall boundary condition with the wall velocity $\bm{u}_w$ as
\begin{equation}
    f_{-j}(\bm{x}_b,t+\Delta t)=f_{j}(\bm{x}_b,t)-2w_j\rho\dfrac{\bm{e}_j\cdot\bm{u}_w}{c_s^2},
\end{equation}
where $w_j$ are the lattice weights and $c_s$ denotes lattice speed of sound. However, this scheme violates both the unitary condition and the positive semidefiniteness of the density matrix, and therefore requires special treatment for implementation on a quantum device in the future work. 

\begin{figure}
    \centering
    \includegraphics[width=0.7\linewidth]{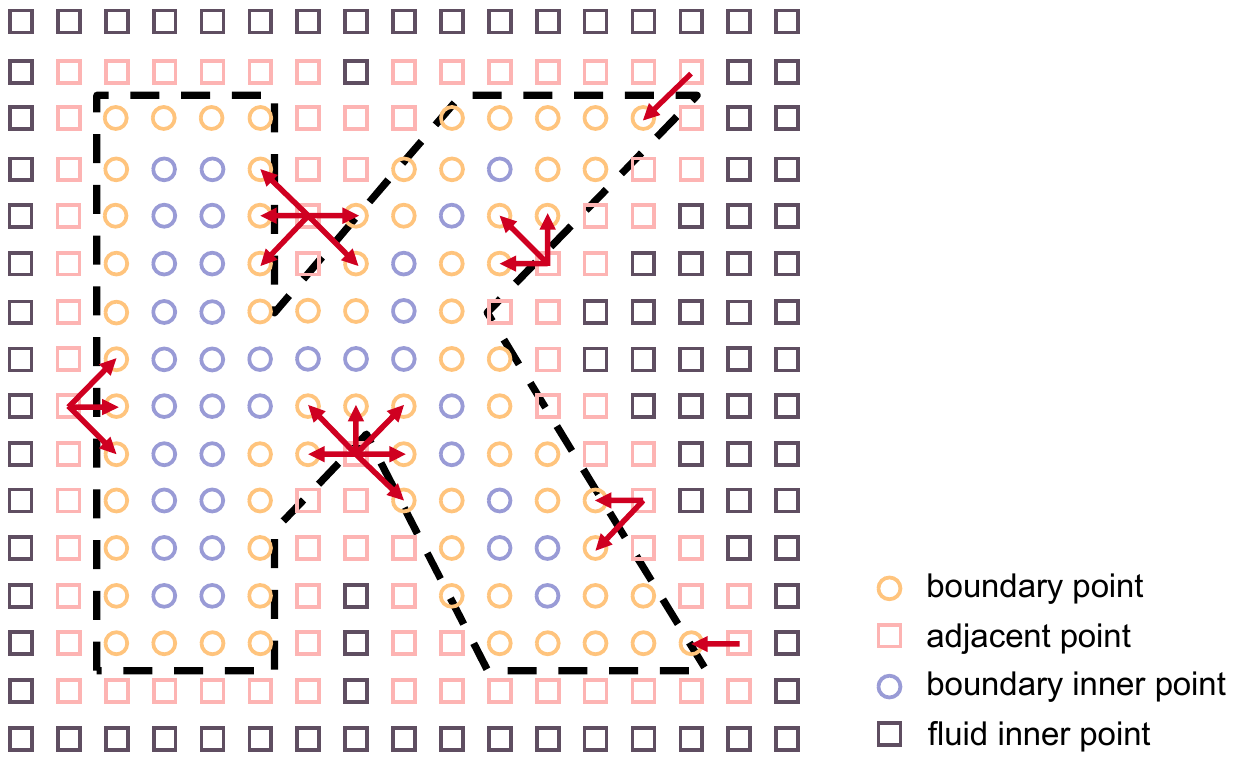}
    \caption{Examples of four types of points near the boundary of a hollow letter K, represented by colored circles and squares. Several bouncing directions of adjacent points under D2Q9 model are indicated by arrows.}
    \label{fig:K_boundary}
\end{figure}

Based on the halfway bounce-back method, we extend the original streaming step and propose a boundary scheme suitable for quantum devices, which takes the form
\begin{equation}\label{eq:quantum_boundary}
    \{\bm{x}, \bm{n}\}\mapsto \dfrac{1}{\rho_{\bm{n}}}\left(\sum_{\mathrm{fluid}}n_j\{\bm{x}+\bm{e}_j\Delta x, \bm{n}\}+\sum_{\mathrm{boundary}}n_j\{\bm{x}, \bm{n}^*\}\right),
\end{equation}
where $\bm{n}^*$ denotes the velocity configuration after bounce-back, and the two summations run over all indices $j$ for which $\bm{e}_j$ points toward the fluid or the boundary, respectively. The particle state $({\bm{x}, \bm{n}})$ is equally divided into $\rho_{\bm{n}}$ branches. Branches whose velocity direction points into the fluid undergo a streaming step to the neighboring grid points, while those pointing toward the boundary are reflected back to the same grid point with their velocity direction reversed. 

To handle boundaries of general geometry on a quantum device, we first classify all grid points into four categories: boundary points $P_{b}=\{\bm{x}|\bm{x}\text{ is in boundary}, \bm{x}\notin P_{bi}\}$, adjacent points $P_{a}=\{\bm{x}|\bm{x}\text{ is in fluid}, \bm{x}\notin P_{fi}\}$, boundary inner points $P_{bi}=\{\bm{x}|\bm{x}+\bm{e}_j\Delta x \text{ is in boundary}, \forall j\}$, and fluid inner points $P_{fi}=\{\bm{x}|\bm{x}+\bm{e}_j\Delta x\text{ is in fluid}, \forall j\}$. In other words, $P_b$ and $P_a$ are the grid points immediately neighboring the boundary. For each adjacent point, the bouncing directions, those directions in which the neighboring grid point is a boundary point, are precomputed and stored during preprocessing. These directions are essential for constructing the quantum circuits for streaming and collision step with boundaries. An example of this classification is provided in Figure~\ref{fig:K_boundary}, where the bouncing directions under the D2Q9 model are marked by arrows. 

Intuitively, boundary points should remain at their positions and should not participate in the streaming step. To enforce this, we must ensure that at every time step, all boundary points have zero probability for any nonzero velocity component. To simultaneously satisfy this condition and implement the boundary scheme of Equation~\eqref{eq:quantum_boundary}, we introduce an additional step ``component exchange'' before the streaming step, as indicated in Figure~\ref{fig:procedure}. 

\begin{figure}
    \centering
    \includegraphics[width=0.9\linewidth]{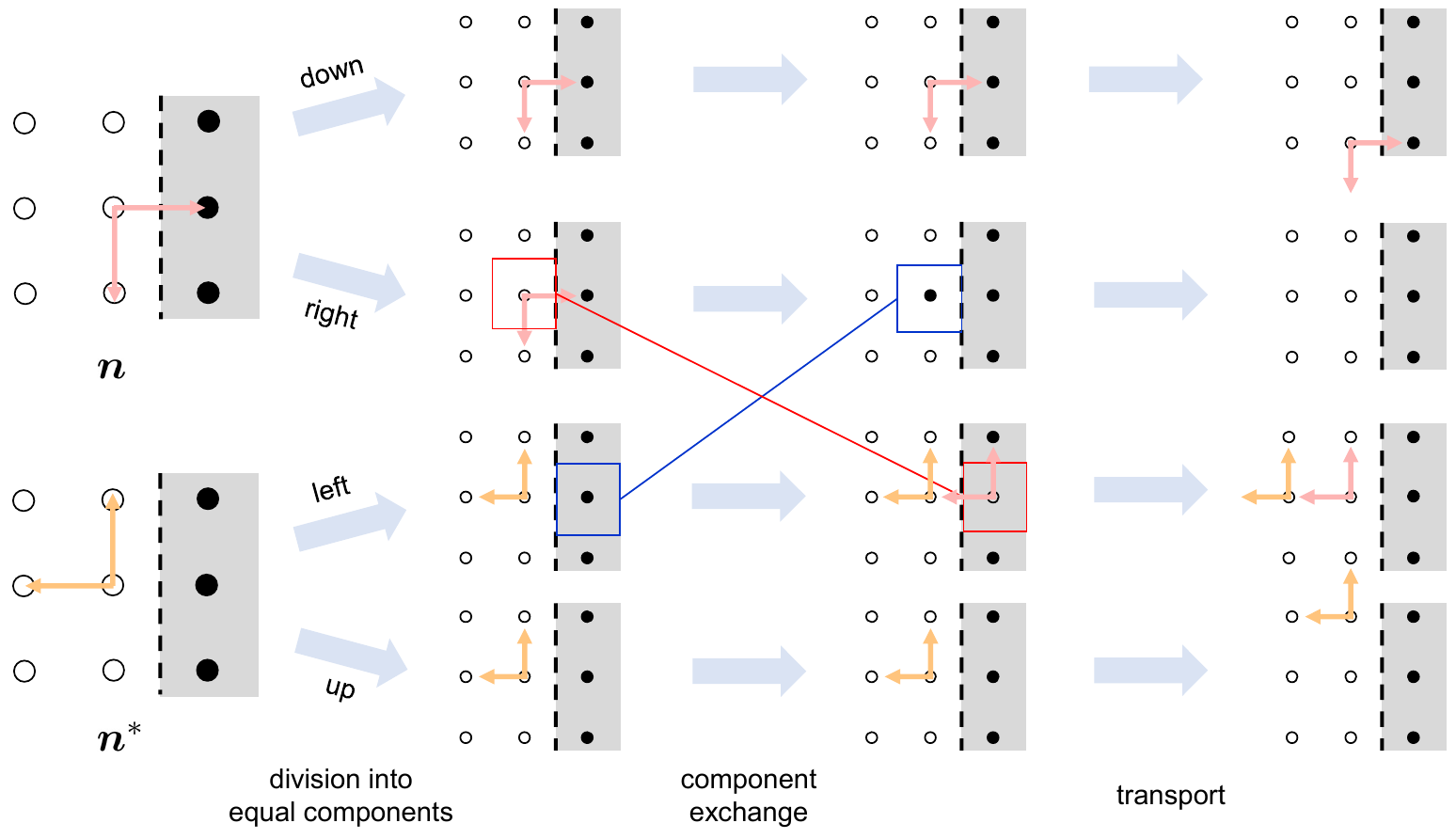}
    \caption{Sketch of the component exchange that implements a QLBM boundary scheme compatible with the streaming step. At the beginning of each time step, every state $\{\bm{x}, \bm{n}\}$ is split equally by velocity configurations via ancilla qubits. The component exchange described in Equation~\eqref{eq:exchange_component} is then performed, as highlighted by the red and purple rectangles. Finally, the streaming step is executed, producing the correct output with boundary treatment as in Equation~\eqref{eq:quantum_boundary}. } 
    \label{fig:boundary_exchange}
\end{figure}

\begin{figure}
    \centering
    \begin{quantikz}
        \lstick{$\ket{\bm{x}}$} &\qw &\gate[wires=3]{\mathrm{SWAP}} &\gate[wires=3]{\mathrm{streaming}} &\qw\\
        \lstick{$\ket{\bm{n}}$} &\qw & & &\qw\\
        \lstick{$\ket{0}$}  &\gate{H}& & &\gate{\mathrm{Trace}}
    \end{quantikz}   
    \caption{Quantum circuit for the streaming step combined with the component exchange operation illustrated in Figure~\ref{fig:boundary_exchange}. The component exchange is indicated by a general SWAP gate. }
    \label{fig:boundary_circuit}
\end{figure}
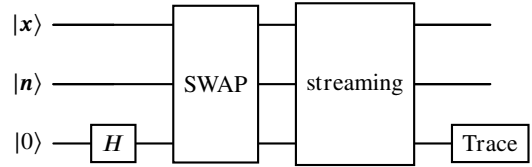

A simplified illustration is provided in Figure~\ref{fig:boundary_exchange}. For a particle state with $\bm{n}=10010$ in the D2Q5 model at an adjacent point, the state is split into two branches: one branch is streamed downward to the neighboring grid point, while the other is bounced back at the same grid point with reversed velocity configuration $\bm{n}^{*}=01100$. This is accomplished by the component exchange operation
\begin{equation}\label{eq:exchange_component}
    \{\bm{x}_{a}, \bm{n}, \bm{e}_j\}\leftrightarrow\{\bm{x}_{a}+\bm{e}_j\Delta x, \bm{n}^{*},\bm{e}_{-j}\},\quad\forall \bm{x}_a\in P_{a},\quad\forall \bm{x}_a+\bm{e}_j\Delta x\in P_{b},
\end{equation}
where the third element $\bm{e}_j$ labels the split branches whose streaming direction points toward the boundary, $\bm{x}_a$ denotes a adjacent point, and $\bm{x}_a+\bm{e}_j\Delta x$ is the neighboring boundary point along the direction $\bm{e}_j$. In Figure~\ref{fig:boundary_exchange}, this exchange is highlighted by the red and purple rectangles, where $\bm{e}_j$ points rightward, and $\bm{e}_{-j}$ points leftward. After the exchange, the streaming step is performed, realizing the boundary scheme of Equation~\eqref{eq:quantum_boundary}. The corresponding quantum circuit is demonstrated in Figure~\ref{fig:boundary_circuit}, where $\lceil\log_2\rho_{\bm{n}}\rceil$ ancilla qubits are introduced to create the equal-weight superposition. The sequence of component exchanges can be regarded as permutations, for which a systematic circuit construction exists, as mentioned earlier. Note that the ancilla qubits are reused from the component exchange step for the streaming step, since both require the same branching structure. The overall circuit does not depend on the specific form of the quantum states, which implies that the it can be designed once and executed repeatedly.

\subsection{Inlet and outlet conditions}
In the LBM, imposing inlet and outlet conditions consists in prescribing the probability distribution over velocity configurations at the inlet and outlet grid points. In quantum computing, this corresponds to directly assigning values to specific amplitudes of the quantum state, which is generally hard to implement as the original state is unknown without measurement. However with the density matrix encoding, this is possible by introducing extra ancilla qubits and applying controlled SWAP operation.

\begin{figure}
    \centering
    \begin{quantikz}
        \lstick{$\ket{\bm{x}}$} &\ctrl{1}&\qw\\
        \lstick{$\ket{\bm{n}}$} &\gate[wires=2]{\mathrm{SWAP}}&\qw\\
        \lstick{$\ket{\bm{n}_t}$}&&\gate{\mathrm{Trace}}
    \end{quantikz}
    \caption{General quantum circuit for implementing inlet and outlet conditions in the QLBM. An ancilla register $\ket{\bm{n}_t}$ is prepared with the prescribed probability distribution over velocity configurations. The SWAP operation is controlled by target grid point $\ket{\bm{x}_t}$, after which the ancilla register is traced out.}
    \label{fig:inlet}
\end{figure}
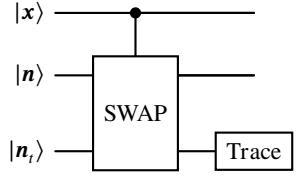

The general circuit for this implementation is illustrated in Figure~\ref{fig:inlet}. An ancilla register $\ket{\bm{n}_t}$ is introduced, prepared with the probability distribution corresponding to prescribed velocity profiles. Conditioned on the position register $\ket{\bm{x}_t}$, a SWAP operation exchanges the velocity register of the target grid point with the ancilla register. The ancilla register is then traced out. For a pure state input of the form $\sum\ket{\bm{n}}\ket{\bm{x}}=A_1\ket{\bm{n}}\ket{\bm{x}_t}+\sum A_i\ket{\bm{n}_i}\ket{\bm{x}_i}$, the circuit executes the transformation
\[\begin{aligned}
    \sum\ket{\bm{n}}\ket{\bm{x}}&\stackrel{\mathrm{ancilla}}\longrightarrow \sum\ket{\bm{n}_t}\ket{\bm{n}}\ket{\bm{x}}\\
    &\stackrel{\mathrm{SWAP}}\longrightarrow A_1\ket{\bm{n}}\ket{\bm{n}_t}\ket{\bm{x}_t}+\sum_i  A_i\ket{\bm{n}_t}\ket{\bm{n}_i}\ket{\bm{x}_i}\\
    &\stackrel{\mathrm{Trace}}\longrightarrow\|A_1\|^2\ket{\bm{n}_t}\ket{\bm{x}_t}\bra{\bm{x}_t}\bra{\bm{n}_t}+\left(\sum_iA_i\ket{\bm{n}_i}\ket{\bm{x}_i}\right)\left(\sum_iA_i^*\bra{\bm{x}_i}\bra{\bm{n}_i}\right).
\end{aligned}\]
The result shows that the circuit transforms a pure state input into mixture of two states. It imposes the prescribed $\ket{\bm{n}_t}$ at the target position $\ket{\bm{x}_t}$ while leaving all other positions unchanged. 

\subsection{Complexity analysis}
We analyze the complexity of our QLBM algorithm with the boundary treatment, evaluated in terms of the number of two-qubit gates and ancilla qubits. We first evaluate the complexity of the component exchange, streaming and collision steps for a single time step. The main part of the component exchange consists of a series of general SWAP gates, each of which requires $O(n)$ two-qubit gates~\cite{Zindorf2025}. Consequently, the component exchange step requires at most $O(qN_{a}n2^q)$ two-qubit gates, where $N_{a}$ is the number of adjacent points. For a particle state with $p$ nonzero velocity components, $\lceil\log_2 p\rceil$ ancilla qubits are required to prepare the equal-weight superposition. This implies that $P\equiv\sum_{p=1}^{q}\lceil\log_2p\rceil$ ancilla qubits are required for this step. The value assignment for inlet and outlet conditions contains $qN_{io}$ multi-controlled SWAP gates, where $N_{io}$ denotes the number of inlet and outlet grid points, and therefore requires $O(qn^2N_{io})$ CNOT gates. In addition, $O(qN_{io})$ ancilla qubits are required for implementation. The streaming step for every velocity configuration contains $O(n)$ multi-controlled CNOT gates, each of which can be implemented with $O(n)$ CNOT gates and $O(n)$ ancilla qubits. Therefore, the total number of CNOT gates for this step is at most $O(2^qn^2)$. As the ancilla qubits used for equal divisions are inherited from the component exchange step, the additional ancilla qubits needed here are the $O(n)$ qubits for implementation of multi-controlled CNOT gates, which can be reused repeatedly. The collision step, in comparison, requires only $O(q)$ CNOT gates and $O(1)$ ancilla qubits, which is negligible compared to the preceding steps. Note that for handling wall boundaries and inlet/outlet conditions, the segment‑wise (SW) decomposition~\cite{Schalkers2024} can be applied to axis‑aligned boundary segments, reducing the numbers $N_a$ and $N_{io}$ actually required in the quantum implementation. For regular boundary shape, this reduction can be significant. 

\begin{figure}
    \centering
    \includegraphics[width=0.8\linewidth]{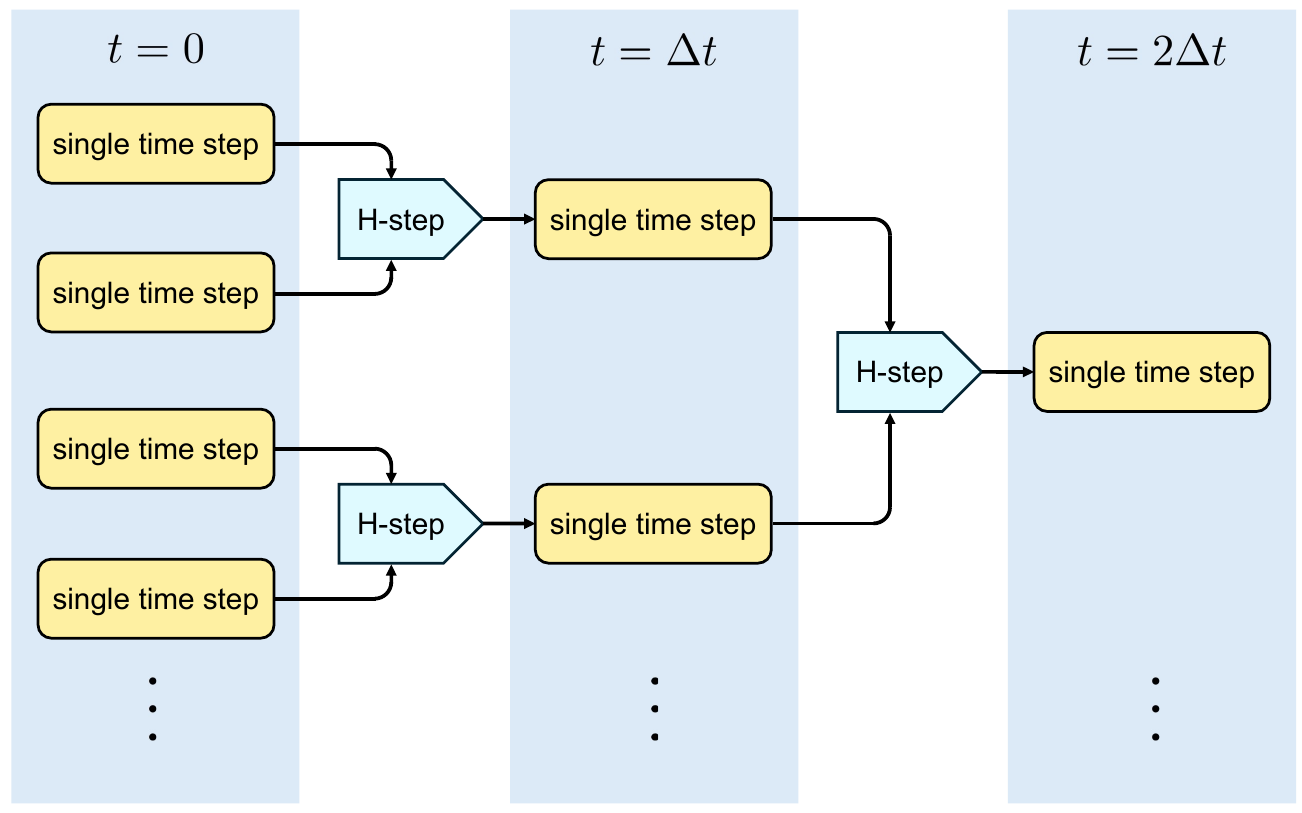}
    \caption{Overall iteration structure of our QLBM algorithm with the H-step. Over the entire simulation, a total of $q^{T/\Delta t}$ copies of the initial states are required. For each single time step, $q$ copies of current quantum states are combined to implement the H-step. The figure illustrates the case $t=2\Delta t$ and $q=2$. }
    \label{fig:complexity}
\end{figure}

Based on the above analysis, we conclude that a single time step requires $O(2^qn^2)$ two-qubit gates and $O(P+n+qN_{io})$ ancilla qubits, where the streaming step being the dominant contribution in single time step. The overall iteration structure for our algorithm is illustrated in Figure~\ref{fig:complexity}. After each time step, the execution of H-step calls for $q$ repetitions of the circuit. This leads to a total of $q^{T/\Delta t}$ repetitions over the entire simulation, where $T$ denotes the simulation time and $\Delta t$ is the time step interval. Consequently, the overall resource requirements of our algorithm can be estimated as $O(q^{T/\Delta t+1}2^qn^2)$ two-qubit gates and $O(q^{T/\Delta t}(P+n+qN_{io}))$ ancilla qubits. 

Thus, the inclusion of boundary and inlet/outlet treatments does not affect the overall gate complexity compared to that in the original QLBM~\cite{Wang2025}, since the streaming step remains the leading term in single time step. 
The speedup ratio and the potential quantum speedup is discussed in Ref.~\cite{Wang2025}. The total complexity of the QLBM is dominated by the H-step part, which requires repeated execution of all other steps. Thus, exploring more efficient algorithm for qubit disentanglement is crucial to improve the algorithm complexity.

\section{Simulation results}\label{sec:result}

\subsection{2D Decaying Poiseuille flow}
We perform several QLBM simulations with the boundary treatment using a quantum emulator with in-house codes on a classic computer. We consider a 2D channel Poiseuille flow in the $x$-$y$ plane with no pressure gradient, where the channel walls are located at $y=\pm 1$. The velocity field is set in the form $\bm{u}=(u(y,t), 0)$, corresponding to a unidirectional flow along the channel. Under these conditions, the NS equation reduces to
\begin{equation}
    \dfrac{\partial u}{\partial t}=\nu\dfrac{\partial^2 u}{\partial y^2},\quad u(y=\pm 1)=0, 
\end{equation}
which includes the no-slip boundary condition. Its characteristic solutions are $u_m(y,t)=\cos(\lambda_m y)e^{-\nu\lambda_m^2t}$ with $\lambda_m=(2m+1)\pi/2$.  

We employ the lowest-order mode $u_0(y,t)=U\cos(\pi y/2)e^{-\nu\pi^2 t/4}$ with $U=0.1$ to test the QLBM with the boundary scheme. The computational domain is $[0,2]\times[-1,1]$, discretized on $128^2$ grid points, which requires a total of $14+9=23$ qubits, where 14 qubits are used to encode the spatial position, and the remaining 9 qubits are needed for the D2Q9 velocity set. Periodic boundary conditions are imposed at $x=0$ and $2$, and the no-slip wall boundary condition is applied at $y=\pm 1$. The corresponding Re for this setup is $\mathrm{Re}=6.37$. The simulation is run up to $t=10$ with $310$ time steps. 

\begin{figure}
    \subfigure{
        \begin{minipage}{0.31\linewidth}
            \centering
        \includegraphics[width=\linewidth]{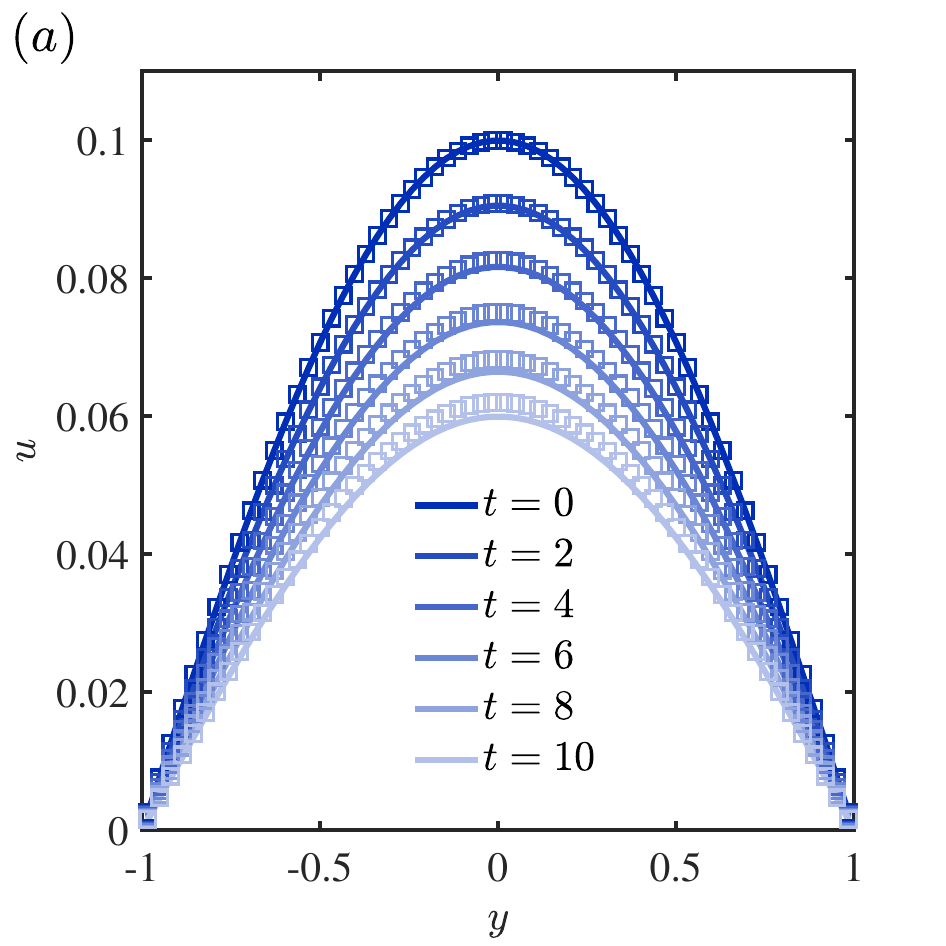}
        \end{minipage}
        \label{fig:Pou_com_general}
    }
    \subfigure{
        \begin{minipage}{0.31\linewidth}
        \centering
        \includegraphics[width=\linewidth]{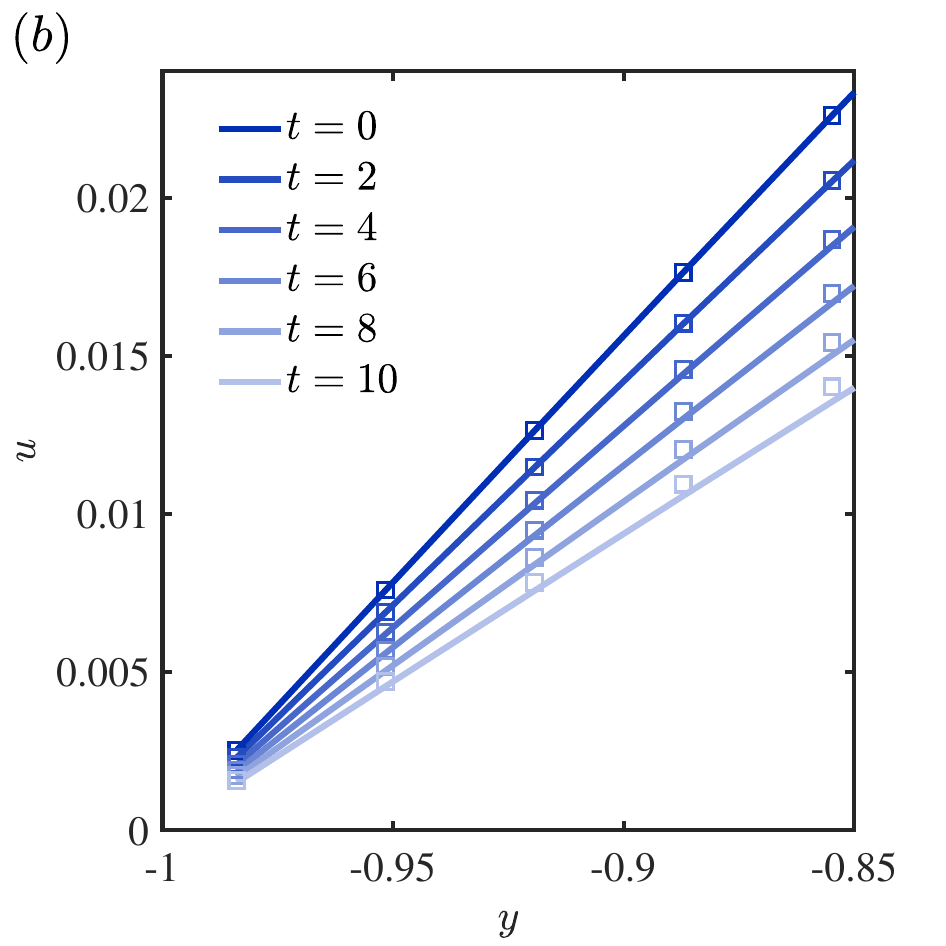}
        \end{minipage}
        \label{fig:Pou_com_boundary}
    }
    \subfigure{
        \begin{minipage}{0.31\linewidth}
        \centering
        \includegraphics[width=\linewidth]{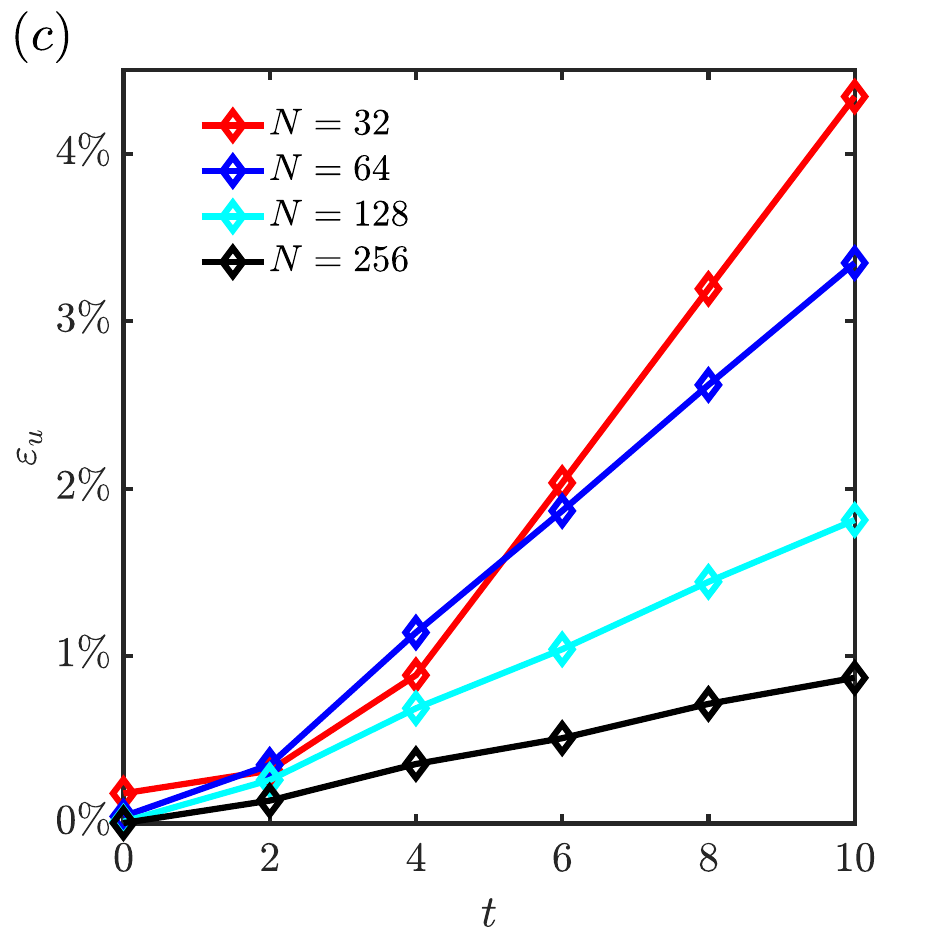}
        \end{minipage}
        \label{fig:Pou_com_error}
    }
    \caption{(a) Comparison of the velocity profile for 2D decaying Poiseuille flow between QLBM results (lines) and analytical solutions (squares) with $N=128$ grid points; (b) comparison near the boundary $y=-1$. (c) Relative error $\varepsilon_u$ for QLBM simulations with $N=32, 64, 128, 256$. }
    \label{fig:Pou}
\end{figure}

The simulation results are demonstrated in Figure~\ref{fig:Pou}. The time-decaying trigonometric profiles are accurately recovered, and the QLBM results agree well with the analytical solutions, as shown in Figure~\ref{fig:Pou_com_general}. In Figure~\ref{fig:Pou_com_boundary}, the velocity at grid points near the wall also matches the analytical predictions closely, indicating that the implemented boundary scheme successfully enforces the no-slip wall condition. Note that with the link-wise boundary scheme, the nearest grid point is located at a distance of $\Delta x/2$ away from the wall. We also perform simulations for this problem using different grid resolutions $N=32, 64, 128, 256$. We evaluate the deviation of the simulation result using the relative L2 error defined as
\begin{equation}\label{eq:L2error}
    \varepsilon_u\equiv \dfrac{\|\bm{u}_{\mathrm{QLBM}}-\bm{u}_{\mathrm{ref}}\|_2}{\|\bm{u}_{\mathrm{ref}}\|_2},
\end{equation}
where $\bm{u}_{\mathrm{QLBM}}$ is the simulation result, $\bm{u}_{\mathrm{ref}}$ is the reference velocity field, and $\|\cdot\|_2$ denotes the standard Frobenius norm for vector and matrix. For the Poiseuille flow case, $\bm{u}_{\mathrm{ref}}$ is taken to be the theoretical solution. The relative error $\varepsilon_u$ for different grid resolutions is illustrated in Figure~\ref{fig:Pou_com_error}. As the grid is refined, the average relative error decreases, and the error grows linearly with the simulation time. This further confirms the accuracy of our QLBM results. 

\subsection{2D backward-facing step flow}
Backward-facing step flow is a well-studied benchmark problem exhibiting rich physics such as flow separation, detachment and recirculating bubbles~\cite{Armaly1983}. The flow structures depend primarily on Re and geometrical parameters, including the step height $h$ and the channel height $H$. The expansion ratio is defined as $H/h$, and the Reynolds number is defined as $\mathrm{Re}=UD/\nu$, where $U$ is the mean inlet velocity, and $D$ is the hydraulic diameter of the inlet channel, which is twice the inlet channel height. At low Re, one of the most important quantities of interest is the length of the recirculation zone. The reattachment point is identified as the location near the bottom wall $y=0$ with zero streamwise velocity. The recirculation length $L_r$ is then defined as the streamwise distance from the step corner to the reattachment point, and it is normalized by $h$.

We set the computational region to $[0,8]\times[0,2]$ with $h=1$ located at $x=0.5$. The domain is discretized on a $512\times 128$ grid. A parabolic velocity profile with a maximum velocity $u_0=0.25$ is imposed at the inlet, while a similar parabolic profile with a maximum velocity $u_0/2$ is prescribed at the outlet. All other boundaries are treated as solid walls. The Reynolds number for this configuration is $\mathrm{Re}=35.5$. The flow field is initialized with zero velocity. 

We also simulate the same problems using OpenFOAM~\cite{Weller1998} and take the steady-state solution as the reference velocity field $\bm{u}_{\mathrm{ref}}$. The computation domain is discretized on a $400\times 100$ grid, with the parabolic velocity profile imposed at the inlet. A pressure outlet is prescribed to be consistent with the parabolic outlet profile, and the viscosity is adjusted to match the target Reynolds number. The governing equations are discretized using the Gauss linear scheme and solved with the Gauss–Seidel method.

\begin{figure}
    \subfigure{
        \begin{minipage}{0.9\linewidth}
            \centering
            \includegraphics[width=\linewidth]{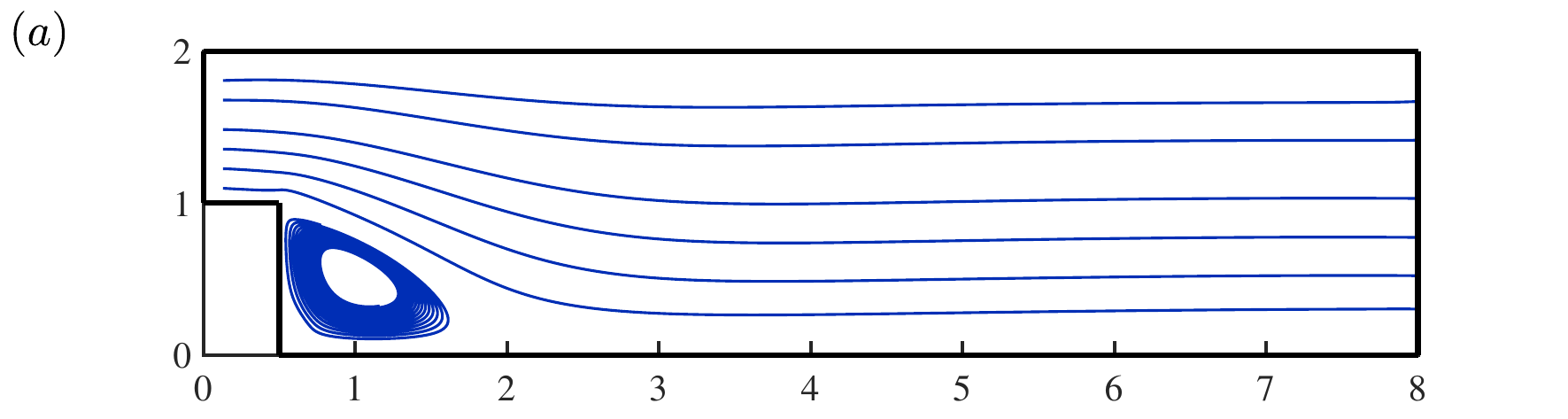}
        \end{minipage}
        \label{fig:backstep_streamline}
    }
    \subfigure{
        \begin{minipage}{0.9\linewidth}
            \centering
            \includegraphics[width=\linewidth]{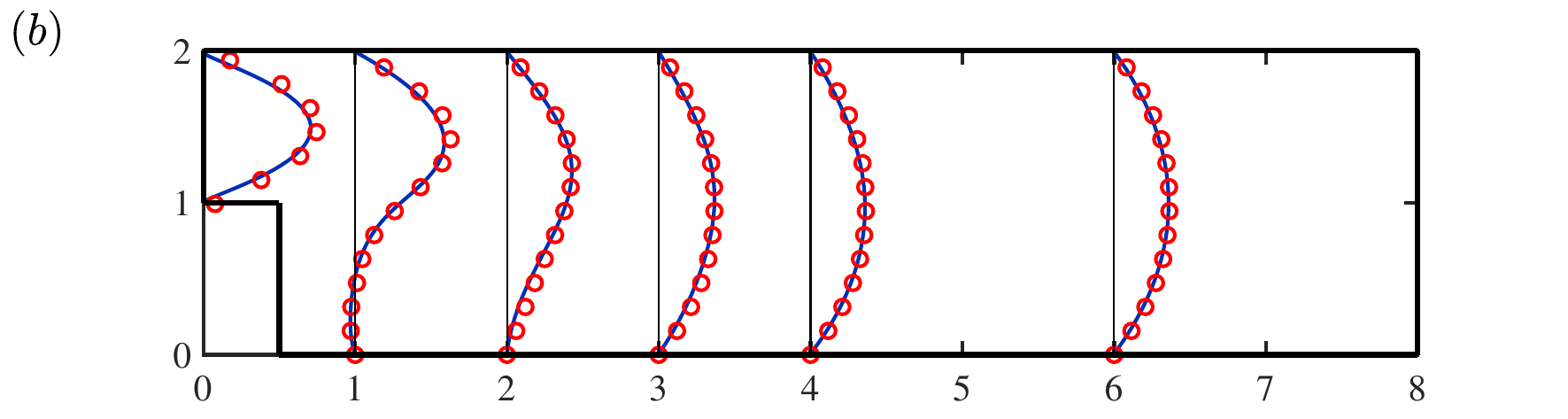}
        \end{minipage}
        \label{fig:backstep_profile}
    }
    \caption{QLBM simulation result for the backward-facing step with $\mathrm{Re}=35.5$ with an expansion ratio of two. (a) Streamlines of the QLBM result, showing the recirculation region with a length of $L_r/h=1.375$. (b) Comparison of velocity profiles between QLBM (blue lines) and OpenFOAM (red circles) results at the inlet and $x=1,2,3,4,6$. The relative error for this simulation is $13.0\%$. }
    \label{fig:backstep}
\end{figure}

The QLBM simulation results are demonstrated in Figure~\ref{fig:backstep}. As shown in Figure~\ref{fig:backstep_streamline}, the recirculation zone behind the step is well captured. The dimensionless recirculation length obtained by QLBM is $L_r/h=1.375$, which agrees closely with the OpenFOAM result of $L_r/h=1.1563$. Figure~\ref{fig:backstep_profile}, compared velocity profiles between the QLBM and OpenFOAM solutions at the inlet and $x=1, 2, 3, 4, 6$. Overall, these results demonstrate that our simulation successfully captures the essential flow features of the backward-facing step problem. The relative error $\varepsilon_u$ for this simulation is $13.0\%$. 

\begin{figure}
    \subfigure{
        \begin{minipage}{0.45\linewidth}
            \centering
            \includegraphics[width=\linewidth]{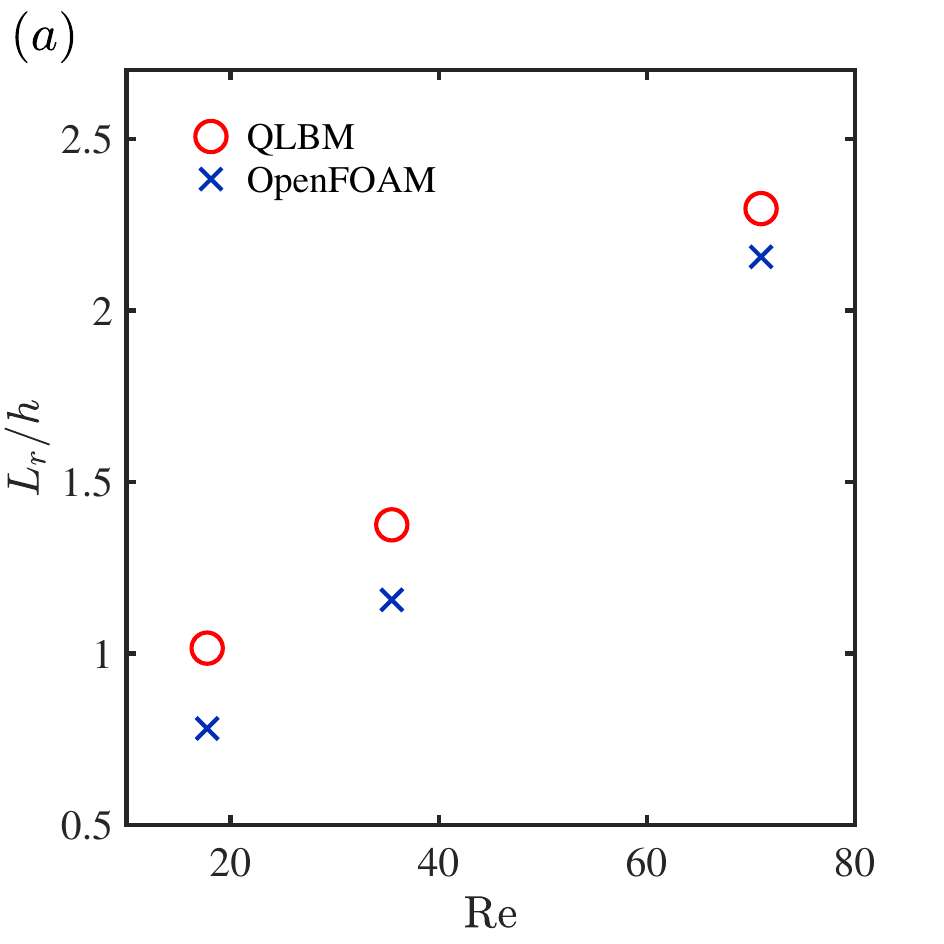}
        \end{minipage}
        \label{fig:backstep_Lr}
    }
    \subfigure{
        \begin{minipage}{0.45\linewidth}
            \centering
            \includegraphics[width=0.92\linewidth]{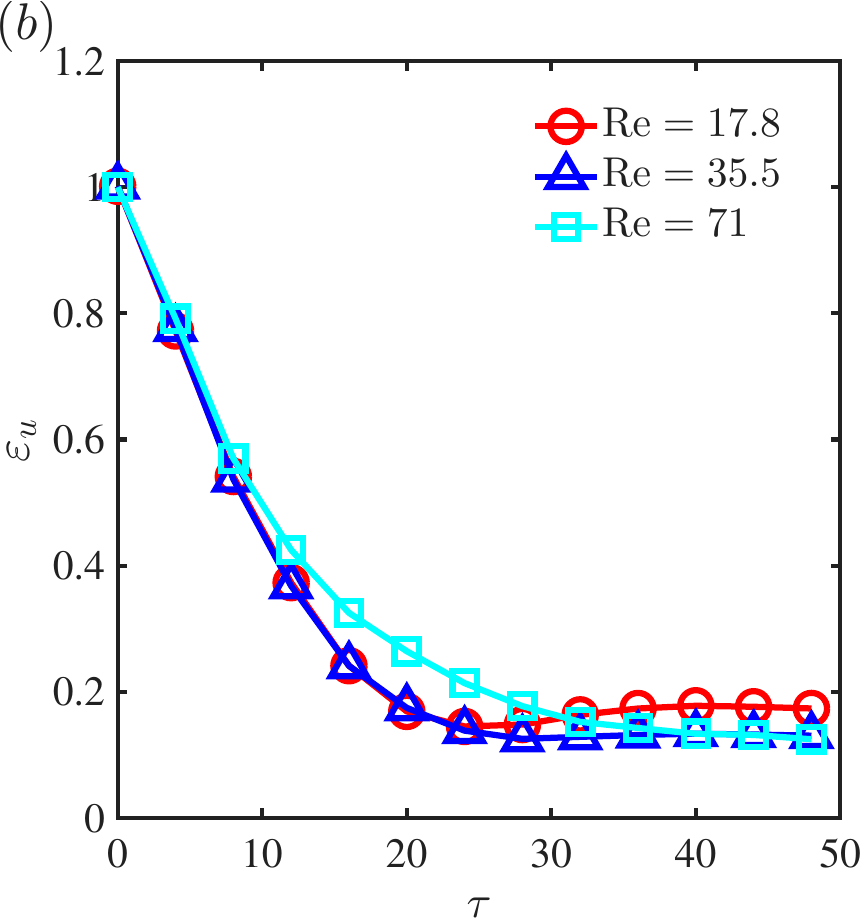}
        \end{minipage}
        \label{fig:backstep_error}
    }
    \caption{QLBM simulation results for the backward-facing step at $\mathrm{Re}=17.8, 35.5, 71$ respectively. (b) Comparison of the normalized recirculation length between QLBM and OpenFOAM simulations across all three Re. Relative errors $\varepsilon_u$ for these cases are $17.0\%$, $13.0\%$, $10.3\%$, respectively. (c) Evolution of $\varepsilon_u$ for three simulations.}
    \label{fig:backstep_collect}
\end{figure}

We also refine the spatial resolution and perform QLBM simulations at various Re, with the results summarized in Figure~\ref{fig:backstep_collect}. The normalized recirculation lengths are compared in Figure~\ref{fig:backstep_Lr}, where the QLBM results are close to the OpenFOAM data, with relative errors of $17.0\%$, $13.0\%$, $10.3\%$, respectively. The evolution of the relative error $\varepsilon_u$ for these three simulations is presented in Figure~\ref{fig:backstep_error}, showing convergence of QLBM simulations at around $\tau\sim 30$, where $\tau$ represents the pseudo-time over which the QLBM simulation converges to a steady solution. The relative error starts at $1$ because the simulation is initialized with a zero velocity field. This further demonstrates the robustness of our algorithm and its reasonable convergence rate. 

\begin{figure}
    \subfigure{
    \begin{minipage}{\linewidth}
        \centering
        \includegraphics[width=0.9\linewidth]{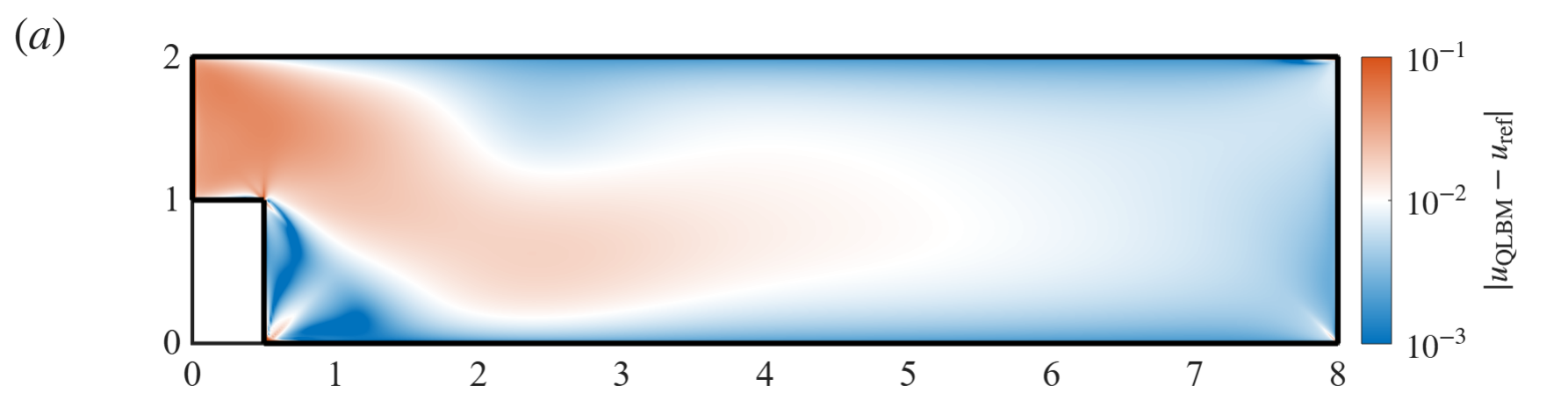}
        \label{fig:backstep_abserror}
    \end{minipage}
    }

    \subfigure{
    \begin{minipage}{\linewidth}
        \centering
        \includegraphics[width=0.9\linewidth]{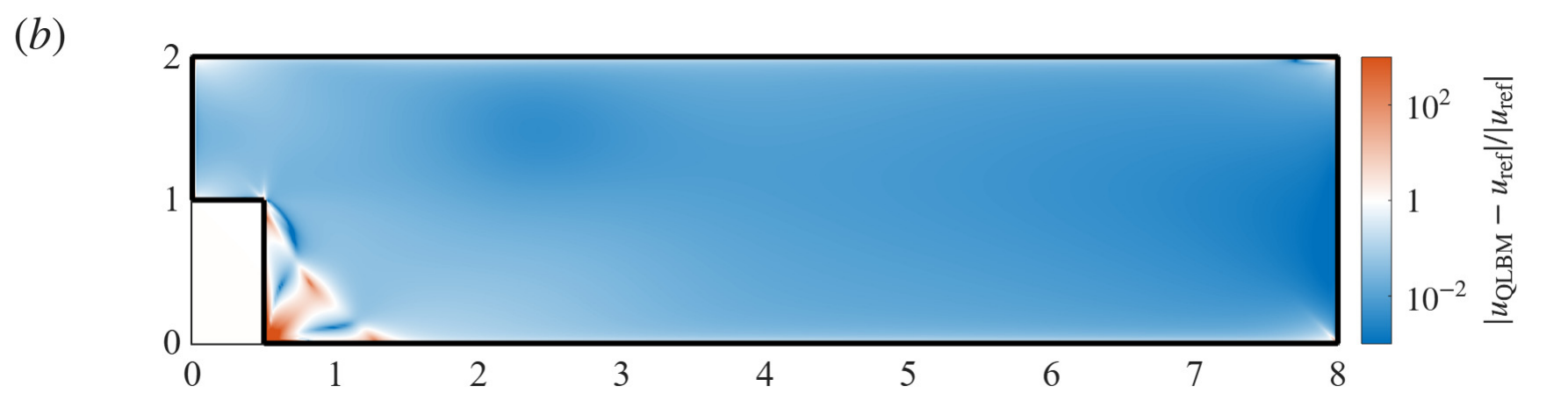}
        \label{fig:backstep_relaerror}
    \end{minipage}
    }
    \caption{Distributions of (a) absolute error and (b) relative error for the backward-facing step flow at $\mathrm{Re}=17.8$. The color scale is logarithmic. }
    \label{fig:backstep_error_distribution}
\end{figure}

We further analyze the deviation of the simulation results in terms of absolute and relative errors, as illustrated in Figure~\ref{fig:backstep_error_distribution}. In Figure~\ref{fig:backstep_abserror}, the high-velocity region originating from the inlet exhibits the largest absolute error, while the corresponding relative error remains low. This suggests that the inlet boundary condition is not perfectly implemented, although it does not significantly affect the overall relative error. The relative error peaks near the corner of the backward-facing step, as shown in Figure~\ref{fig:backstep_relaerror}, where the absolute velocity magnitude is small. This indicates that the error mainly originates from the implementation of the inlet/outlet conditions and the limited accuracy of the wall boundary treatment. The inlet condition is enforced through direct value assignment and may deviate after the streaming and collision steps. Furthermore, the wall boundary treatment may not be sufficiently accurate to resolve fine details near complex boundaries where the velocity magnitude is small.

\subsection{Flow past a obstacle}
Next we test our QLBM on the flow past a cylinder. Unlike the previous simulations, this setup involves a curved boundary and thus represents a broader class of boundary conditions. The computational domain is chosen as $[0, 8]\times[0, 2]$. The cylinder, with a diameter of $D_c=0.4$, is located at $x=0.5, y=1$. Parabolic velocity profiles with a peak velocity $U=0.25$ are imposed on the inlet and outlet, and no-slip wall boundary conditions are applied at $y=0, 2$. The Reynolds number for this scenario is defined as $\mathrm{Re}=UD_c/\nu$. Two simulations are performed using different grid resolutions: a $256\times 64$ grid with $\mathrm{Re}=6.6$, and a $1024\times 256$ grid with $\mathrm{Re}=43$. $\bm{u}_{\mathrm{ref}}$ are taken as steady-state solutions from OpenFOAM simulations. The simulation setup is similar to that used for the backward-facing step flow.

\begin{figure}
    \subfigure{
    \begin{minipage}{0.5\linewidth}
        \centering
        \includegraphics[width=\linewidth]{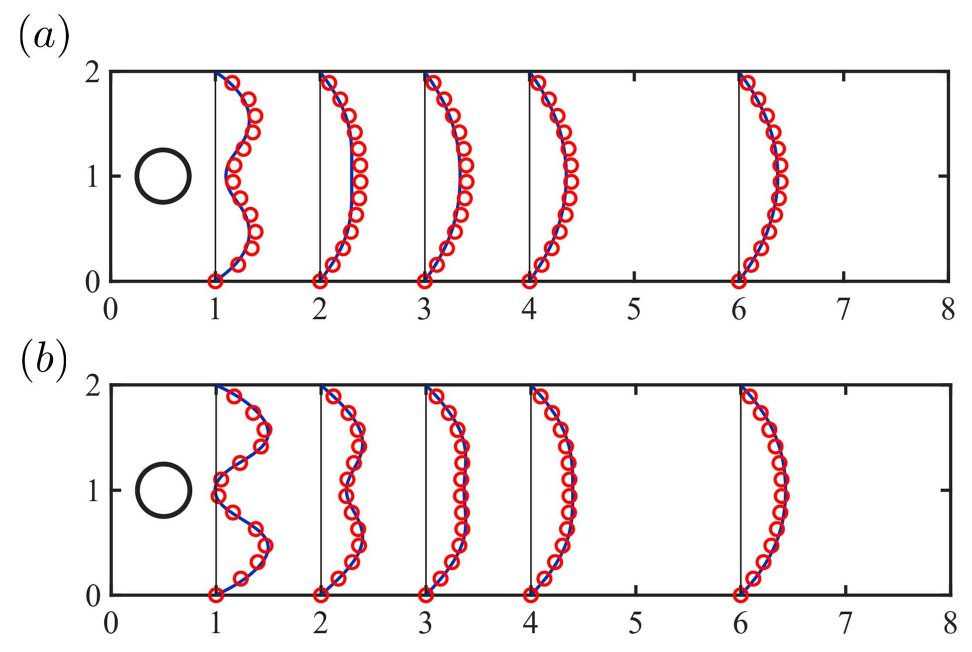}
    \end{minipage}
    }
    \subfigure{
    \begin{minipage}{0.45\linewidth}
        \centering
        \includegraphics[width=0.75\linewidth]{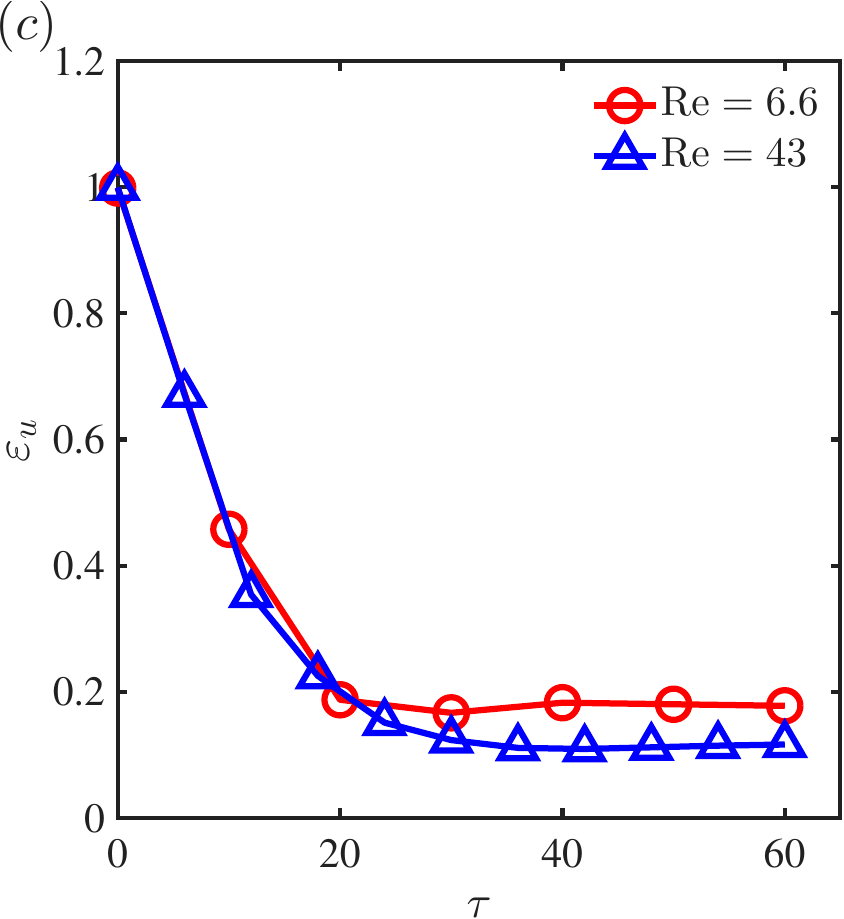}
    \end{minipage}
    }
    
    \caption{QLBM simulation result for flow past a cylinder with (a) $\mathrm{Re}=6.6$ and (b) $\mathrm{Re}=43$ respectively. Comparisons of velocity profiles with corresponding OpenFOAM simulations at $x=1, 2, 3, 4, 6$ are presented. Blue lines represent the QLBM results, and red circles represent the OpenFOAM results. The relative errors are $17.34\%$ and $12.53\%$ respectively. (c) Evolution of $\varepsilon_u$ for two simulations. }
    \label{fig:cylinder}
\end{figure}

The results for the two cases are illustrated in Figure~\ref{fig:cylinder}. Comparisons of the velocity profiles between the QLBM results and OpenFOAM simulations at locations $x=1, 2, 3, 4, 6$ are shown in Figures~\ref{fig:cylinder}(a) and (b). The relative errors are $17.34\%$ and $12.53\%$ for these two cases. It can be seen that the low-velocity region behind the cylinder is well captured by our algorithm, and the velocity profiles from our simulation matches closely with the OpenFOAM results. The evolution of $\varepsilon_u$ is shown in Figure~\ref{fig:cylinder}, indicating convergence at around $\tau\sim 40$. These results indicate that our algorithm continues to deliver accurate results even for boundaries with more complex geometries. 

\begin{figure}
    \centering
    \includegraphics[width=0.9\linewidth]{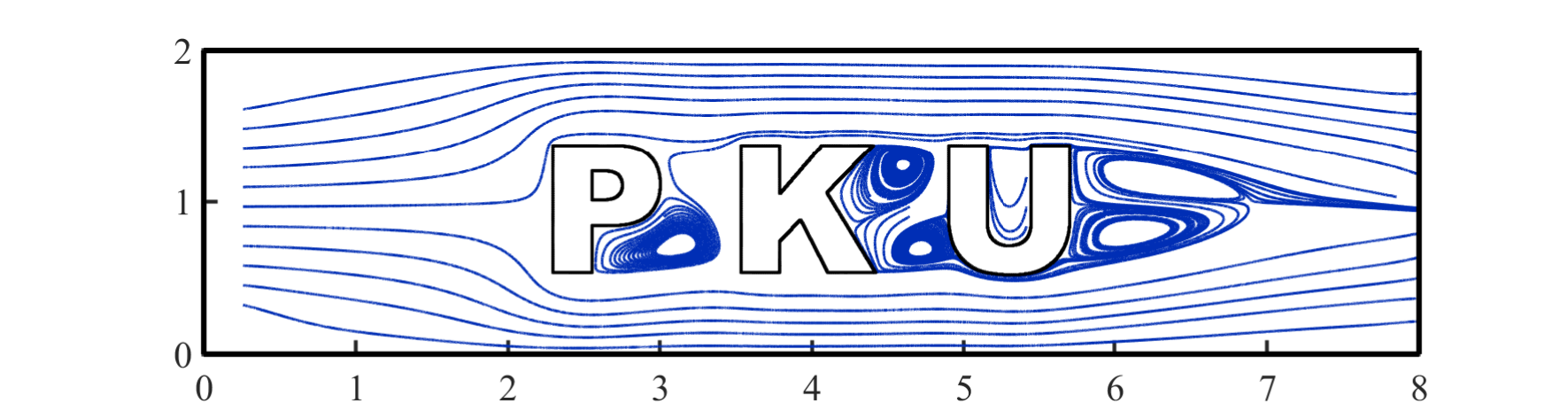}
    \caption{Streamlines in the QLBM simulation for a flow past the letters ``PKU'' with $\mathrm{Re}=64.8$.}
    \label{fig:PKU}
\end{figure}

We also simulate a flow past the letters ``PKU''. The computational domain and the boundary conditions are kept the same as before, with the center of the letters located at $x=4, y=1$ and their overall size being $L_x=2.5$ and $L_y=0.61$. A $2048\times 512$ grid is applied, with $\mathrm{Re}=U L_y/\nu=64.8$. The QLBM results in Figure~\ref{fig:PKU} show that the streamlines pass around the complex letter shapes, with a series of vortices captured in the gaps between the letters and behind them. This result further demonstrates the capability of our algorithm in handling boundaries with complex geometries and capturing rich flow phenomena. 

\section{Conclusions}\label{sec:conclusion}
We have proposed a QLBM algorithm with boundary treatment capable of handling the wall boundary condition of arbitrary geometry, while also accommodating inlet and outlet velocity conditions. 
Building upon the QLBM~\cite{Wang2025} which combines LGCA and LBM, we introduce a more general density-matrix encoding scheme that affords greater flexibility in quantum circuit design. We show that all QLBM operations can be implemented with this encoding. Within the Kraus operator theory, we re-interpret the quantum circuit originally proposed in Ref.~\cite{Wang2025}. 

For wall boundary conditions, we design a quantum algorithm and its corresponding circuit implementation based on the half-way bounce-back scheme. A carefully designed component exchange step is introduced, allowing the streaming step to naturally enforce the no-slip condition without requiring any modification to the streaming circuit itself. The correctness and efficacy of our algorithm are validated on a series of 2D benchmark flows, ranging from canonical test cases to more complex geometries, demonstrating the accuracy of our method, along with the generality and extensibility of our boundary treatment.

Several limitations of the current method warrant further investigation. First, the H‑step introduces considerable complexity, which may undermine the quantum speedup for moderate‑sized problems. Second, handling boundaries of arbitrary geometry may incur additional overhead, both in terms of circuit depth and the number of ancillary qubits required. Third, moving boundary problems remain entirely unaddressed in the present framework. These issues collectively pose significant challenges for practical implementation on near‑term quantum hardware. In future work, we plan to explore simplified realizations of the H‑step, or alternative formulations that altogether bypass this step. We also aim to extend our boundary treatment to more general configurations, including time‑dependent and deformable boundaries, with the ultimate goal of applying our methodology to industrially relevant flow problems.

\section*{Acknowledgement}
This work has been supported by the National Natural Science Foundation of China (grant nos.~12525201, 12432010, and 12588201). 

\section*{Data availability}
The open source code release is provided in \url{https://github.com/YYgroup/QLBM_wallBC}.

\bibliographystyle{elsarticle-num-names}
\bibliography{cas-refs}

\end{document}